\documentclass[twocolumn,10pt]{article}

\usepackage[T1]{fontenc}
\usepackage[a4paper,margin=18mm,columnsep=7mm]{geometry}
\usepackage{ifpdf}
\ifpdf
  \usepackage{graphicx}
\else
  \usepackage[dvipdfmx]{graphicx}
\fi
\usepackage{amsmath,amssymb}
\usepackage{booktabs}
\usepackage{multirow}
\usepackage{url}

\usepackage{algorithm}
\usepackage{algpseudocode}
\newcommand{\codebreakunderscore}{\textunderscore\penalty100\hskip0pt\relax}
\newcommand{\code}[1]{\texttt{\let\_\codebreakunderscore #1}}
\newcommand{\freal}{\code{cu\_freal}}
\newcommand{\fcomplex}{\code{cu\_fcomplex}}
\newcommand{\PB}{\ensuremath{p}}

\newcommand{\ulp}{\textsc{ulp}}
\newcommand{\rndn}{\textsc{rndn}}

\usepackage{listings}
\newcommand{\doi}[1]{DOI: \url{#1}}

\begin{document}

\title{Construction and Performance Evaluation of an Arbitrary-Precision Floating-Point Arithmetic Environment on CUDA}
\author{Tomonori Kouya\thanks{Faculty of Science and Engineering, Otemon Gakuin University, Japan}}

\twocolumn[
\begin{@twocolumnfalse}
\maketitle
\begin{abstract}
Arbitrary-precision floating-point arithmetic is used to solve ill-conditioned problems in scientific
computing, and MPFR and MPC are the de facto standard libraries on CPUs. The multiple-precision libraries
currently available for CUDA GPUs do not combine MPFR-compatible rounding, complex arithmetic, and elementary
functions that can be called inside a CUDA kernel. We present mpc\_cuda, which consists of two layers:
(i) a port of the mini-GMP-based MPFR and MPC to CUDA device code, produced by rerunnable source-conversion
scripts, with a per-thread bump arena for temporary storage and a per-thread copy of the exponent range and
exception flags, and (ii) compile-time fixed-precision types cu\_freal/cu\_fcomplex whose addition,
subtraction, and multiplication agree bitwise with MPFR round-to-nearest. In 3.3 million test cases covering
all rounding modes, mixed precisions, special values, and the limits of the exponent range, executed
concurrently on the GPU, no discrepancy from the system MPFR/MPC was found on either machine used: in the
3.02 million cases with the five deterministic rounding modes, the values, ternary values, and flags
returned by layer (i) were identical, and in the 0.28 million cases with faithful rounding every value was
one of the two admissible ones; the same tests exposed two defects of the conversion, which
we report. Performance was measured on an NVIDIA GB10 (20 CPU cores) and an H100 (32 CPU cores). At 1024
bits, layer (i) evaluates elementary functions 17--31$\times$ (GB10) and 14--77$\times$ (H100) faster than
the same code on one CPU core, but not faster than the GMP-based system libraries on all CPU cores of the
same machine; it is 2.2--6.9$\times$ and 4.2--5.9$\times$ faster than the latter for AXPY, and for matrix
multiplication on a par with them on GB10 and up to 2.5$\times$ faster on H100. For AXPY, layer (ii) is
the fastest of the GPU libraries measured (CUMP, CAMPARY, and layer (i)) up to 2048 bits on both machines.
In simultaneous polynomial root-finding by a higher-order Ehrlich--Aberth method, with identical inputs,
identical iteration counts, and comparable forward errors, layer (ii) obtains the roots 1.6--11.8$\times$
faster than the fastest multithreaded CPU configuration for degrees 64 to 192 on GB10 and
1.1--5.7$\times$ faster for degrees 96 to 192 on H100, and is not consistently faster at lower degrees.
From these results, we derive guidelines for selecting an implementation by required semantics, precision,
and problem size.
\end{abstract}
\begin{center}
\end{center}
\bigskip
\end{@twocolumnfalse}
]
{\renewcommand{\thefootnote}{\fnsymbol{footnote}}%
\footnotetext[1]{Faculty of Science and Engineering, Otemon Gakuin University, Japan}}

\section{Introduction}
\label{sec:introduction}

Many problems in scientific computing are too ill-conditioned for binary64 arithmetic, and the number of
cases that require multiple-precision floating-point arithmetic with a longer significand is increasing.
On CPUs, MPFR~\cite{mpfr,mpfr_paper} and MPC~\cite{mpc} are the de facto standard: they provide real and
complex arithmetic and elementary functions at a precision chosen at run time, with correct rounding in every
directed and to-nearest rounding mode (and a faithful mode with weaker guarantees), on top of the multiple-precision natural-number layer (MPN) of GNU MP (GMP)~\cite{gmp}.
Numerical software that relies on these semantics cannot be moved to a GPU unless the same semantics are
available there.

Several multiple-precision libraries exist for CUDA~\cite{cuda} GPUs, but each covers only a part of this
functionality (Section~\ref{sec:related}). GQD~\cite{gqd} is limited to double-double and quad-double
precision. CAMPARY~\cite{campary_paper2016} supports longer multi-component expansions but provides neither
correct rounding nor elementary functions, and its cost grows rapidly with the number of components.
CUMP~\cite{cump} ports GMP's \code{mpf} layer and therefore provides only addition, subtraction, and
multiplication without a rounding guarantee. MPRES-BLAS~\cite{mpres_blas_jpdc,mpres_blas} provides
multiple-precision BLAS routines based on a residue number system; its scalar addition and multiplication can
be called inside a kernel, but it provides neither MPFR/MPC rounding semantics nor division or elementary
functions on the device. To the best of our knowledge, no publicly available library provides
MPFR/MPC-compatible real and complex arithmetic with elementary functions that can be called inside a CUDA
kernel.

Porting MPFR itself to CUDA is the direct way to fill this gap, but MPFR depends on GMP's
architecture-specific MPN kernels and on dynamic memory allocation, and we previously abandoned an attempt to
extend CUMP toward MPFR because of the complexity of its division. Two observations make the port feasible.
First, GMP ships mini-GMP, a single-file portable MPN/MPZ implementation, and the current MPFR can be built on
mini-GMP alone; MPC is a thin layer on MPFR. mini-GMP uses only schoolbook algorithms, which is acceptable
for the precisions of a few hundred to a few thousand bits that are typical in scientific computing.
Second, the remaining work is a large amount of mechanical rewriting, which can be automated by source
conversion scripts.

This paper presents mpc\_cuda~\cite{mpc_cuda}, an open-source library built along these lines, and makes the
following contributions.
\begin{enumerate}
  \item \textbf{A port of MPFR/MPC to CUDA device code that preserves upstream semantics.} Conversion
        scripts turn the unmodified mini-GMP, MPFR, and MPC sources into device-callable code
        (Section~\ref{sec:port}); no arithmetic routine is replaced. Beyond the mechanical conversion, the
        port manages temporary memory and MPFR's global state (exponent range, flags) separately for each
        GPU thread (Section~\ref{sec:arena}). We also report the defects of the conversion that a
        full-significand comparison with the system libraries revealed (Section~\ref{sec:div}).
  \item \textbf{Compile-time fixed-precision types.} \freal{} and \fcomplex{} keep the limb-array
        representation of MPFR but fix the number of limbs at compile time, which removes dynamic
        allocation from the arithmetic (Section~\ref{sec:fixed}). Their addition, subtraction, and
        multiplication agree bitwise with MPFR/MPC round-to-nearest.
  \item \textbf{An evaluation that identifies where each layer is effective.} We verify the numerical
        results against MPFR/MPC over the full significand, including ternary values and flags under
        concurrent execution (Section~\ref{sec:verification}); measure arithmetic kernels against the
        GMP-based system MPFR/MPC on all CPU cores and against CUMP, CAMPARY, and MPRES-BLAS
        (Section~\ref{sec:benchmarktest}); apply both layers to simultaneous polynomial root-finding and
        compare the time to solution at comparable accuracy (Section~\ref{sec:rootfinding}); and summarize
        which implementation to select for given semantics, precision, and problem size, including the
        cases in which the GPU is slower than the CPU (Section~\ref{sec:discussion}).
\end{enumerate}
This e-print is a thoroughly revised version of its first version (v1): the library was corrected, all
experiments were repeated under stricter conditions, and several claims of v1 are withdrawn or corrected.
``The previous version'' below refers to v1, and the differences are stated where they matter.

A generative AI coding assistant was used during development; because it is a development tool and not a
technical contribution of this paper, its role is described only in Appendix~\ref{sec:appendix_ai}.

\section{Related work and positioning}
\label{sec:related}

\subsection{Classification of software multiple-precision arithmetic}

Software multiple-precision floating-point arithmetic can be classified along two independent axes: how the
significand is represented, and when the precision is determined.

\paragraph{Representation of the significand}
The \textbf{multi-component approach} represents a number as an unevaluated sum of $k$ binary64 (or binary32)
numbers and builds its arithmetic on error-free transformations. Double-double (DD) and
quad-double (QD)~\cite{qd} are representative; GQD~\cite{gqd} is their GPU counterpart, mX\_real~\cite{mx_real}
provides double- to quad-word types as C++ templates, and CAMPARY~\cite{campary_paper2016} generalizes the
approach to $k$ components. This approach is fast when $k$ is small, the precision can be selected only in
steps of one component, and the results are generally not correctly rounded. In addition, because every
component is a binary64 number, the low-order components of a long expansion underflow, which bounds the
attainable accuracy (Section~\ref{sec:cump}).

The \textbf{integer limb-array approach} holds the significand as an array of machine integers (limbs) and
allows the precision to be selected in bits. GMP's \code{mpf}, MPFR, and MPC belong to this class, as does
ARPREC and its GPU port GARPREC~\cite{gqd}, which use arrays of floating-point words holding integer digits.
On GPUs, CUMP~\cite{cump} ports \code{mpf}. The IEEE binary128 format, which CUDA supports in device code since version 12.8 on GPUs of compute
capability 10.0 or higher~\cite{cuda_fp128} (hence on GB10 but not on H100), can be regarded as the fixed
113-bit member of this class.

The \textbf{residue-number-system (RNS) approach} represents the significand by its residues modulo a set of
small moduli, which makes addition and multiplication of the significand digit-parallel.
MPRES-BLAS~\cite{mpres_blas_jpdc,mpres_blas} uses this representation and provides multiple-precision BLAS
routines for GPUs.

\paragraph{When the precision is determined}
In MPFR, CUMP, and GARPREC the precision is a run-time property of each object, so the storage for the
significand must be allocated dynamically. In QD, GQD, mX\_real, CAMPARY, and binary128 the precision is
part of the type and is fixed at compile time, so a value is a plain fixed-size object.

\subsection{Position of this work}

Table~\ref{tab:libs} summarizes the libraries above together with the two layers of mpc\_cuda. The
run-time-precision layer (\code{cu\_mpfr}/\code{cu\_mpc}) belongs to the limb-array class with run-time
precision; unlike CUMP, it is the MPFR/MPC code itself and therefore provides all rounding modes, complex
numbers, and elementary functions. The fixed-precision layer (\freal/\fcomplex) is not a new representation:
it is the limb-array representation with the precision fixed at compile time, a combination that none of the
libraries in Table~\ref{tab:libs} provides beyond 113 bits. Its aim is to obtain the plain-object property of
the multi-component libraries while keeping MPFR's rounding for the basic operations.

\begin{table*}[t]
\centering
\footnotesize
\caption{GPU-capable multiple-precision libraries and the two layers of mpc\_cuda.
``Kernel'' indicates whether individual operations can be called inside a user-written CUDA kernel.
The entries for the third-party libraries were taken from their documentation and source code
(GQD: README of its repository; mX\_real, CAMPARY, CUMP, MPRES-BLAS~1.7.0: source code; binary128: CUDA
documentation~\cite{cuda_fp128}).}
\label{tab:libs}
\setlength{\tabcolsep}{3pt}
\begin{tabular}{@{}p{0.13\textwidth}p{0.15\textwidth}p{0.17\textwidth}p{0.17\textwidth}p{0.21\textwidth}p{0.09\textwidth}@{}}
\toprule
library & representation & precision & rounding & operations on GPU & kernel \\
\midrule
GQD~\cite{gqd}            & multi-component (binary64) & 106 / 212 bits, compile time & not correctly rounded & arithmetic, $\sqrt{\ }$, elementary functions & yes \\
mX\_real~\cite{mx_real}   & multi-component (binary32/64) & 2--4 words, compile time & not correctly rounded & $+,-,\times,\div,\sqrt{\ }$ & yes \\
CAMPARY~\cite{campary_paper2016} & multi-component (binary64) & $k$ components, compile time & not correctly rounded & $+,-,\times,\div,\sqrt{\ }$ & yes \\
binary128~\cite{cuda_fp128} & IEEE 754 & 113 bits, compile time & IEEE 754 for $+,-,\times,\div$; functions $\le 1$~\ulp & arithmetic, $\sqrt{\ }$, elementary functions & yes ($\ge$\,\code{sm\_100}) \\
GARPREC~\cite{gqd}        & digit array & run time & not correctly rounded & (not released for current GPUs) & --- \\
CUMP~\cite{cump}          & limb array (GMP \code{mpf}) & run time (64-bit limb units) & truncation, no guarantee & $+,-,\times$ & yes \\
MPRES-BLAS~\cite{mpres_blas_jpdc,mpres_blas} & RNS + binary exponent & compile time (moduli set; 120--2400 bits predefined) & not correctly rounded & BLAS levels 1--3, SpMV; scalar $+,-,\times$ & yes ($+,-,\times$) \\
\midrule
\code{cu\_mpfr}/\code{cu\_mpc} & limb array (MPFR) & run time (bit units) & all MPFR/MPC modes & real/complex arithmetic, elementary functions & yes \\
\freal/\fcomplex          & limb array & compile time (32-bit units) & \rndn{} for $+,-,\times$; not correctly rounded otherwise & real/complex arithmetic, elementary functions & yes \\
\bottomrule
\end{tabular}
\end{table*}

\paragraph{Libraries compared in this paper}
In Section~\ref{sec:cump} we compare mpc\_cuda with CUMP, CAMPARY, and MPRES-BLAS, the libraries in
Table~\ref{tab:libs} that support precisions of several hundred bits or more on current GPUs.
For MPRES-BLAS the comparison is restricted by how its precision is defined: the precision is fixed at
compile time by a set of RNS moduli, the predefined sets provide 120, 240, 360, 480, 720, 960, 1200, \dots,
2400 bits and none of them equals 128, 256, 512, 1024, or 2048 bits, and its interface consists of BLAS-level
host functions, each of which launches several kernels. We therefore report it separately
(Section~\ref{sec:mpres}) and do not rank it against the other libraries at equal precision.
GARPREC is excluded because no release for current GPUs exists: the maintained repository of its
authors contains only GQD, and the only copy of the GARPREC sources that we could obtain (bundled with
the test suite of MPRES-BLAS) does not compile with CUDA~13, since it depends on the CPU library ARPREC and on
CUDA interfaces that have been removed (\code{cudaThreadSynchronize} and fields of \code{cudaDeviceProp}).
GQD, mX\_real, and binary128 are limited to 212 bits or less and are not compared. Consequently, where this
paper states that an mpc\_cuda type is the fastest at 128 bits, the statement is relative to the libraries
measured here; it does not extend to the multi-component libraries, which are the natural choice at 212
bits or less when their rounding properties are acceptable.

\section{Design and implementation of mpc\_cuda}
\label{sec:transport_cuda}

\subsection{Two layers and their numerical contracts}
\label{sec:layers}

mpc\_cuda provides two kinds of types (Table~\ref{tab:layers}). \code{cu\_mpfr\_t}/\code{cu\_mpc\_t} are MPFR
and MPC ported to the GPU: the precision is a run-time property of each object, and the API is the upstream
API with the prefix \code{cu\_}, which allows the port to coexist with the system \code{<mpfr.h>} and
\code{<mpc.h>} in one program. \code{cu\_freal<\PB>}/\code{cu\_fcomplex<\PB>} fix the precision \PB{} at
compile time and are written as ordinary values (Listing~\ref{lst:usage}).

The two layers make different guarantees, and we state them separately throughout this paper. For the
run-time-precision layer, the intended contract is that of upstream MPFR/MPC: the same value for every
operation, precision, and rounding mode. For the fixed-precision layer, addition, subtraction, and
multiplication return the same value as MPFR/MPC round-to-nearest at \PB{} bits, whereas division and
elementary functions are computed with guard bits and carry no rounding guarantee. The evidence for each
contract is given in Section~\ref{sec:verification}.

\begin{table}[htb]
\centering
\scriptsize
\setlength{\tabcolsep}{2pt}
\caption{The two layers of mpc\_cuda and their numerical contracts.}
\label{tab:layers}
\begin{tabular}{@{}lp{0.34\columnwidth}p{0.38\columnwidth}@{}}
\toprule
 & \textbf{Run-time precision} & \textbf{Fixed precision}\\
 & \code{cu\_mpfr\_t}, \code{cu\_mpc\_t} & \code{cu\_freal<\PB>}, \code{cu\_~fcomplex<\PB>}\\
\midrule
Precision  & run time, per object, in bits & compile-time constant \PB{} (multiple of 32)\\
Storage    & limb array in device heap or per-thread arena & $\lceil \PB/64\rceil$ limbs inside the object\\
Rounding   & all MPFR/MPC modes (argument) & \rndn{} only (no argument)\\
$+,-,\times$ & upstream MPFR/MPC code & same value as MPFR \rndn{} / \code{MPC\_RNDNN}\\
Division   & upstream MPFR/MPC code (see Appendix~\ref{sec:appendix_tie}) & Newton iteration with guard bits, not correctly rounded\\
Elem.\ func. & upstream MPFR/MPC code (correctly rounded) & 128 guard bits, no guarantee (observed errors: \S\ref{sec:verification})\\
Exceptions & $\pm0$, $\pm\infty$, NaN, flags, ternary value, exponent range & none (no $\infty$/NaN, no flags, unchecked exponent)\\
\bottomrule
\end{tabular}
\end{table}

\begin{lstlisting}[language=C++,basicstyle=\scriptsize\ttfamily,numbers=none,caption={Example use of the fixed-precision layer (256-bit AXPY kernel)},label={lst:usage},float=t]
#include "cu_freal.cuh"
using cu_fp::cu_freal;
__global__ void axpy(int n, cu_freal<256> a,
        const cu_freal<256>* x, cu_freal<256>* y) {
  int i = blockIdx.x*blockDim.x + threadIdx.x;
  for (; i < n; i += gridDim.x*blockDim.x)
    y[i] = a * x[i] + y[i];   // no dynamic allocation
}
\end{lstlisting}

\subsection{Porting by source conversion scripts}
\label{sec:port}

We adopted mini-GMP (\code{gmp-6.3.0/mini-gmp}) as the basis for the port. With a single C source, mini-GMP
provides the multiple-precision integers \code{mpz} and the limb-sequence layer \code{mpn} (64-bit limbs).
MPFR~4.2.2 can be built on mini-GMP with the \code{--with-mini-gmp} option, the missing \code{mpn} routines
being supplied by \code{mpfr-mini-gmp.c} shipped with MPFR. MPC~1.4.1 is a thin complex layer on top of MPFR.

Rewriting the upstream sources by hand and keeping a fork would make it difficult to follow upstream updates.
We therefore leave the upstream sources untouched and apply the GPU-oriented modifications with conversion
scripts (\code{tools/cudafy\_*.py}). \code{cudafy\_minigmp.py} reads the pristine \code{mini-gmp.\{c,h\}} and
generates CUDA sources in which every function reachable from device code is qualified with
\code{\_\_host\_\_ \_\_device\_\_}. Functions that can run only on the CPU (those that use stream I/O or memory
reallocation) are identified by scanning the source token by token and are left host-only. Similarly,
\code{cudafy\_mpfr.py} converts 263 MPFR sources and \code{cudafy\_mpc.py} converts 90 MPC sources; a further
script renames the exported symbols with the \code{cu\_} prefix. Because the original structure is preserved,
a new upstream release is followed by rerunning the scripts.

We chose conversion over a rewrite for two reasons. The correctness of MPFR (significand and exponent
handling, special values, and all rounding modes) has been established over a long period~\cite{mpfr_paper},
and reproducing it from scratch is impractical; and scripting the conversion keeps the cost of following
upstream low. Note, however, that reusing the upstream code does not by itself guarantee the correctness of
the port: the code is compiled by a different compiler, the conversion itself can be defective
(Section~\ref{sec:div}), and the code is executed by many threads concurrently (Section~\ref{sec:arena}). The properties inherited from
upstream and those that must be established separately are distinguished in Section~\ref{sec:verification}.

The CUDA compiler nvcc compiles C++ and does not accept several C99/C11 features used by the upstream
sources. The adaptations applied by the scripts are as follows.
\begin{enumerate}
  \item \textbf{C11 \code{\_Noreturn}}: not accepted; \code{-DMPFR\_HAVE\_NORETURN} is dropped.
  \item \textbf{Inline assembly}: the x86\_64-specific assembly in \code{mpfr-longlong.h}
        (\code{umul\_ppmm}, etc.) cannot be used on the GPU; \code{-DNO\_ASM} selects the C fallback.
  \item \textbf{\code{realloc}}: not available in device code; it is emulated by allocate, copy, and free
        (\code{mg\_cuda\_realloc}).
  \item \textbf{C99 \code{\_Complex}}: not available in device code; the corresponding switch in
        \code{mpc.h} is disabled and \code{mpc\_set\_dc}/\code{get\_dc} are removed.
  \item \textbf{\code{gmp.h} shim}: a bridging header maps MPC's \code{gmp.h} to mini-GMP, with empty
        dummies for the types that mini-GMP lacks.
  \item \textbf{Macro token pasting}: \code{set\_x\_x.c} generates setters from one macro and thereby
        references functions absent from mini-GMP (\code{mpfr\_set\_f}/\code{\_q}); trapping stubs are
        provided for them.
\end{enumerate}
Formatted I/O and string conversion remain host-only, and the special functions of MPFR (gamma, zeta, erf,
Bessel, Airy, etc.) are not included in the current port.

\subsection{Division, and defects of the conversion found by verification}
\label{sec:div}

\paragraph{Division}
In the library version used for the previous version of this paper, the device build replaced MPFR's general division
\code{mpfr\_div}~\cite{HarveyZimmermann,MollerGranlund} by a routine built from \code{mpz} operations,
because the device build of the upstream routine had returned \code{inf} for precisions of three or more limbs
at an early stage of the port. The previous version of this paper attributed this to a miscompilation by nvcc. We
withdraw that statement, for which we had no evidence: at the time, lowering the optimization level
(\code{-O0}, \code{-G}) had not changed the behavior, and with the current conversion scripts the
\emph{unmodified} upstream \code{mpfr\_div}, compiled for the device by the same compiler (nvcc~13.0.88),
agrees in sign, exponent, and every bit of the significand with the system MPFR in all division cases of the
verification of Section~\ref{sec:verification}. We could not reproduce the original failure and do not know
which of the corrections made to the scripts since then removed it. The run-time-precision layer now uses
the upstream division on the device as well as on the host, so that it contains no replaced arithmetic
routine. The \code{mpz}-based routine can still be selected when the library is built
(\code{-DMPC\_CUDA\_MPZ\_DIV}); it is not used in this paper.

\paragraph{Defects found by the full-significand tests}
The tests of Section~\ref{sec:verification} exposed two defects of the conversion that were present in the
version used for the previous version of this paper. They had gone unnoticed because the earlier tests compared
results after conversion to binary64 and only in the default exponent range.
\begin{enumerate}
  \item \textbf{Internal functions left host-only.} The scripts qualify every prototype and definition
        with \code{\_\_host\_\_ \_\_device\_\_}. The prototypes of the internal functions
        \code{mpfr\_overflow} and \code{mpfr\_underflow} begin with an attribute macro and were missed, as
        were several \code{static} helper functions of MPC whose definitions are formatted differently from
        the rest (among them the Taylor-series branch of \code{mpc\_asin}). nvcc reports a call from device
        code to a host-only function (diagnostic 20011) but compiles the program, and that diagnostic had
        been suppressed in the build because it is also issued, harmlessly, for the stream I/O in functions
        that are never called on the device. On the device such a call has no effect. As a consequence, the
        device code never signalled overflow or underflow: a result whose exponent was outside the current
        range was returned as an ordinary number and no flag was raised.
  \item \textbf{Unprefixed symbols.} A few external symbols of MPC (\code{mpcr\_*}, \code{set\_pi\_over\_2},
        \code{sqrt\_int64}) had not been renamed with the \code{cu\_} prefix and collided with the system
        MPC when both were linked statically.
\end{enumerate}
Both are corrected in the scripts. The library is now also compiled once with diagnostic 20011 enabled, and
the remaining instances are checked to refer only to stream I/O, string, and assertion functions in
host-only code. The experiments of the previous version of this paper stayed far from the limits of the default
exponent range, which is why the first defect was not visible in them; all experiments of this paper were
rerun with the corrected library. The episode shows that reusing the upstream sources does not by itself
transfer their correctness: only the properties that are tested on the device can be claimed.

\subsection{Per-thread arena and per-thread state}
\label{sec:arena}

\begin{figure*}[t]
\centering
\includegraphics[width=.8\textwidth]{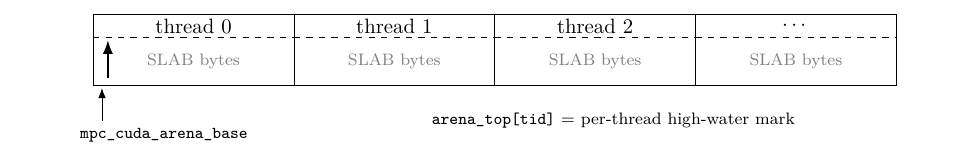}
\caption{Per-thread bump arena. One \code{cudaMalloc} block is divided into fixed-size slabs, one per launched
thread; allocation advances the offset, free is a no-op, and reset returns the offset to 0.}
\label{fig:arena}
\end{figure*}

\paragraph{Temporary memory}
The MPFR and MPC routines allocate working memory internally: a single call to \code{cu\_mpfr\_div} or
\code{cu\_mpc\_mul} allocates and frees many temporaries along its call chain. On the CPU this is inexpensive,
but the device-side \code{malloc}/\code{free} of CUDA is a single allocator shared by all threads of the
device, and the allocation requests of thousands of threads are serialized.

mpc\_cuda therefore serves these allocations from a per-thread \emph{bump arena}. One contiguous block
obtained by \code{cudaMalloc} is divided into fixed-size partitions (\emph{slabs}), one per launched thread,
indexed by the global thread number (Fig.~\ref{fig:arena}). The allocator supports three operations.
\begin{itemize}
  \item \textbf{Allocate}: add the requested size to the thread's current offset
        \code{mpc\_cuda\_arena\_top[tid]} and return the previous position. No search and no lock is
        involved, and a thread accesses only its own slab.
  \item \textbf{Free}: no operation.
  \item \textbf{Reset}: set the thread's offset to zero, which releases the whole slab at once.
\end{itemize}
Because Free does nothing, everything allocated between two resets stays alive; the arena is a scratch area
for one unit of work, and a kernel calls \code{mpc\_cuda\_arena\_reset()} before moving to the next unit.
Values that must survive a reset, such as the accumulator of an inner product, are placed on the thread's
stack (\code{MPFR\_DECL\_INIT}). The hooks are common to mini-GMP and, through \code{mpfr-gmp.c}, to MPFR and
MPC, so the arena serves all three layers. If no arena is installed, or if a unit of work does not fit in its
slab, the allocation falls back to the device \code{malloc}; an undersized slab therefore reduces performance
but does not change the results.

The total size of the arena is (number of launched threads) $\times$ (slab size). The slab size is set to
the peak usage of one unit of work, which can be measured directly as the value of
\code{mpc\_cuda\_arena\_top[tid]} after one unit of work without a reset. The requirement is approximately
proportional to the precision; at 1024 bits, the real AXPY and inner product use 32\,KB slabs and the complex
ones 256\,KB slabs. The launch parameters used in each experiment are listed in
Appendix~\ref{sec:appendix_repro}.

\paragraph{Memory layout and access pattern}
With this layout, the slabs of the threads of one warp are separated by the slab size, so that when the
threads of a warp execute the same instruction on their own temporaries, the accessed addresses are strided
by 32\,KB or more rather than adjacent. Such accesses cannot be coalesced. An interleaved layout, in which
the $j$-th limb of all threads is stored contiguously, would allow coalescing but is incompatible with the
pointer-based interface of the upstream code, which assumes that each significand is a contiguous array.
We adopt the slab layout for this reason; its cost is quantified in Section~\ref{sec:analysis}.

\paragraph{Exponent range and flags}
Upstream MPFR keeps the exponent range (\code{emin}, \code{emax}), the exception flags, and the default
precision and rounding mode in global variables, which are thread-local in a thread-safe build. In the
version of mpc\_cuda used for the previous version of this paper, these variables were shared by all GPU threads. This
is a data race: the flags set by one thread are visible to the others, and the routines that temporarily
widen the exponent range (\code{MPFR\_SAVE\_EXPO\_MARK}/\code{FREE}, used by most functions other than the
basic operations) expose the modified range to concurrently running threads and restore a range that
another thread may have saved in the meantime.

In the current version each thread has its own \emph{state block} with these five variables. The conversion
script replaces their declarations by macros that expand to a field of the block returned by an inline
function. On the host the block is a thread-local object. On the device it is an element of an array in
device memory, indexed by the same global thread number as the arena; the array is allocated by
\code{mpc\_cuda\_tstate\_install($n$)} for the $n$ launched threads and initialized from the state of the
calling host thread, so that a range or default set on the host before the launch is inherited by every
GPU thread. Because the upstream code accesses the variables only through their names, no other source
change is needed: \code{MPFR\_SAVE\_EXPO\_MARK}/\code{FREE}, \code{mpfr\_set\_emin}, and the flag functions
read and write only the block of the calling thread. A program can read back the blocks after a kernel
(\code{mpc\_cuda\_tstate\_download}) to obtain the flags raised by each thread. The arena and the state blocks are indexed by the thread number within one kernel launch. They therefore
separate the threads of one kernel; two kernels that use the library concurrently on the same device (for
example from different streams) would share slabs and blocks, and this case is neither supported nor
tested. If no blocks are installed,
the device threads share a single block as before; all experiments in this paper install them.

Two further objects of the upstream code are shared between threads and mutable. The caches of the constants
$\pi$, $\log 2$, Euler's constant, and Catalan's constant had already been bypassed in the previous
version: each call recomputes the constant at the requested precision, on the host and on the device.
The cache of the overflow and underflow thresholds in \code{mpfr\_exp}, a set of \code{static} variables,
had been overlooked; the script now turns it into local variables that are computed in each call. We
inspected the remaining file-scope objects of the converted sources that are reachable from device code;
they are read-only tables. The cost of the additional indirection is measured in
Section~\ref{sec:analysis}.

\subsection{Fixed-precision types}
\label{sec:fixed}

\code{cu\_freal<\PB>} is a plain structure consisting of a sign, an exponent, and
$N=\lceil \PB/64\rceil$ 64-bit limbs, where \PB{} is a template parameter and a multiple of 32. It uses the
same significand layout as MPFR, $x = \pm M\cdot 2^{e-64N}$, so that finite nonzero values can be converted exactly to and
from an \code{mpfr\_t} of the same precision. The type is header-only and can also be compiled for the host, where the limb
products use \code{\_\_int128}.

The type is deliberately minimal and does not provide the exceptional values of MPFR. The exponent is a
64-bit signed integer without range checks: there is no overflow or underflow handling, and the
operations assume that all exponents, including intermediate sums of exponents, remain representable; the
behavior is undefined otherwise. Zero is represented by a reserved exponent value and
carries no meaningful sign. There are no infinities and no NaN, no exception flags, and no ternary return
values; division by zero and arguments outside the domain of a function (e.g., the logarithm of a
non-positive number) return unspecified finite values. Programs that need any of these must use the
run-time-precision layer.

Because $N$ is a compile-time constant, all loops over limbs have constant trip counts, the temporaries are
fixed-size local arrays, and no operation allocates memory. The compiler is therefore free to keep the limbs
in registers or in the thread's local memory; neither the arena nor the device heap is used. Section~\ref{sec:analysis} reports how many registers and how much local memory the compiler actually
uses at each precision.

\paragraph{Addition, subtraction, and multiplication}
Multiplication rounds the exact $2N$-limb product once to \PB{} bits with round-to-nearest-even; addition
and subtraction align the operands, keep the round and sticky information of the shifted-out part, and
round once. Each operation thus returns the correctly rounded
result, which is the value returned by \code{mpfr\_mul}/\code{mpfr\_add}/\code{mpfr\_sub} with
\code{MPFR\_RNDN} at \PB{} bits. For \code{cu\_fcomplex<\PB>}, each component of a product,
$\mathrm{re}=\circ(a_r b_r-a_i b_i)$ and $\mathrm{im}=\circ(a_r b_i+a_i b_r)$, is computed from the exact
$2N$-limb partial products and rounded once, which matches \code{mpc\_mul} with \code{MPC\_RNDNN}. The cost of
a multiplication is $O(N^2)$; no subquadratic method is used.

\paragraph{Truncated product with an exactness test}
On the device, for real multiplication with $4\le N\le32$ (256 to 2048 bits), the product is first computed without its low-order
columns; the device code for complex multiplication does not use this truncation and always forms the full
$2N$-limb products. Write the significands of the operands as $a=\sum_{i=0}^{N-1}a_i\beta^i$,
$b=\sum_{j=0}^{N-1}b_j\beta^j$ with $\beta=2^{64}$, and let $T$ be
the exact sum of the partial products $a_ib_j\beta^{i+j}$ with $i+j\ge N-3$, which costs about half of the
$N^2$ limb products. The omitted part $D=ab-T$ consists of the $(N-3)(N-2)/2$ partial products with
$i+j\le N-4$; since each is smaller than $\beta^{i+j+2}$ and column $k$ contains $k+1$ of them,
$0\le D<(N-2)\,\beta^{N-2}$, which includes all carries because it bounds the omitted \emph{value}. After
the normalizing shift by zero or one bit, the truncated product is split into the $N$ leading limbs $H$,
the word $R$ below them, and the next word $G$, whose unit is
$\beta^{N-2}$ or $\beta^{N-2}/2$; the words of the exact product in the same positions are $H^\ast$,
$R^\ast$, $G^\ast$. By the bound, the exact value exceeds the truncated one by less than $2N$ units of $G$.
Hence, if $G<\beta-2N$, no carry propagates out of $G$, so that $H=H^\ast$ and $R=R^\ast$, and
$G\le G^\ast$. The \PB-bit result consists of the leading \PB{} bits of $H$; the round bit is the bit that
follows them and the sticky bit is the disjunction of all bits below it. When $\PB=64N$ the round bit is
the leading bit of $R$; when \PB{} is not a multiple of 64 (288, 544, or 1056 bits, for example), the
$s=64N-\PB$ low bits of $H$ lie below the result, the round bit is bit $s-1$ of the lowest limb of $H$, and
the bits of $H$ below it and all of $R$ belong to the sticky part. The same code path is used in both
cases; only these bit positions, which are compile-time constants, differ. In either case the round bit
and all sticky bits down to the end of $R$ are exact, and the only unknown is whether the part below $R$
vanishes. This matters only for the \emph{half-way pattern}, which occurs when the round bit is one and
all known bits below it are zero: the lower 63 bits of $R$ when $\PB=64N$, or the bits of $H$ below the
round bit together with all bits of $R$ when $\PB<64N$. If $G\neq0$, the part below $R$ is nonzero and the
sticky bit is one; if $G=0$ and the pattern is not half-way, the decision does not depend on it. The
implementation therefore uses the truncated product unless
\[
  G\ge\beta-2N \quad\text{or}\quad (G=0 \text{ and half-way pattern}),
\]
and in these two cases recomputes the full $2N$-limb product. The result is thus that of the exact product
for all inputs, including exact ties, which always take the second branch. For operands with uniformly
distributed significands the full product is needed with a probability of about $2N/2^{64}$
($3\times10^{-18}$ at 2048 bits); it is needed regularly only for sparse operands whose product is short.
We did not count the fallbacks in the benchmarks. The 27 million test cases of
Section~\ref{sec:verification} include constructed exact ties and structured significands, which exercise
both branches. On GB10, disabling the truncated product (\code{-DCU\_FP\_DEV\_MULHIGH\_MAX\_N=3}) increases
the time of the AXPY of Section~\ref{sec:cump} by 7\%, 28\%, 10\%, and 3\% at 256, 512, 1024, and 2048
bits (0.0190 to 0.0202, 0.0348 to 0.0447, 0.177 to 0.194, and 0.444 to 0.455\,ms; mean of 20 launches);
its contribution to the times of Table~\ref{tab:cump} is thus moderate.

\paragraph{Division and elementary functions}
Division, square root, cube root, and the elementary functions (real: \code{exp}, \code{expm1}, \code{log},
\code{log1p}, \code{sin}, \code{cos}, \code{tan}, \code{atan}, \code{sinh}, \code{cosh}, \code{asin},
\code{acos}, \code{atanh}, \code{pow}; complex: division, \code{sqrt}, \code{exp}, \code{log}, \code{sin},
\code{cos}, \code{tan}, \code{sinh}, \code{cosh}) are evaluated with the correctly rounded basic operations
at a working precision of exactly $\PB+128$ bits (Newton iteration for the reciprocal and the roots,
argument reduction and series for the functions), and the result is rounded once to \PB{} bits. They carry
no rounding guarantee. For the functions, a guarantee would require Ziv's strategy of recomputing at
increasing precision, whose working precision is not bounded at compile time, and with a fixed number of
guard bits the accuracy is lost wherever the evaluation cancels (Section~\ref{sec:verification} shows where).
Division is a different case: it could be rounded correctly at the cost of one additional multiplication
and comparison (a remainder check), which the current implementation does not perform. Programs that require
the MPFR/MPC semantics of division or of the functions should use the run-time-precision layer, subject to
the upstream division issue documented in Appendix~\ref{sec:appendix_tie}.

\section{Numerical verification}
\label{sec:verification}

This section states what is verified, by which method, and over which range, separately for the two layers.
We use ``bit-identical'' only in the following sense: the sign, the exponent, and the complete \PB-bit
significand of the GPU result are equal to those of the reference computed by an independently built
system MPFR/MPC on the CPU. Agreement after conversion to binary64, agreement of iteration counts, or
agreement of printed residuals is not sufficient evidence for this statement and is not used as such.

\subsection{Method}

\paragraph{Fixed-precision layer}
For addition, subtraction, and multiplication of \freal{} and \fcomplex, the results computed on the device
are converted exactly to \code{mpfr\_t} and compared with the results of the system MPFR/MPC
(\code{MPFR\_RNDN}, \code{MPC\_RNDNN}) by \code{mpfr\_equal\_p}, for 15 precisions from 32 to 2048 bits
(the multiples of 64 and values such as 96, 160, 288, 544, and 1056 that leave a partial limb).
The inputs are random significands with exponent differences of up to 40 and of up to $\PB+200$, and corner
cases: structured significands, exponent gaps from 0 to 66 and near \PB, almost complete cancellation, exact
ties of products, and zero operands. Because the type has no signed zero (Section~\ref{sec:fixed}), the
sign of a zero result is not compared.

For division, square root, cube root, and the functions, the device results at 64, 128, 256, and 512 bits
are compared with a reference computed by MPFR/MPC at $\PB+192$ bits, and the error is expressed in units
of the last place (\ulp) of the \PB-bit result. Each device result is also compared with this reference
rounded to \PB{} bits, called the \emph{rounded reference} below, and agreement with it is reported. The
rounded reference is the correctly rounded \PB-bit value unless the
exact result lies within a relative distance of about $2^{-\PB-192}$ of a rounding boundary without lying
on it; such double-rounding effects are not rigorously excluded by this test. Each of the 26 operations is evaluated on 100,000 inputs
per precision in each of four classes (41,600,000 inputs); an input is counted only if the reference result
is a finite number, which excludes the large arguments for which \code{exp}, \code{expm1}, \code{sinh}, or
\code{cosh} overflows and leaves 41,256,099 results, 20,456,099 of them in the first group below. The
classes are: arguments of order one, tiny arguments, large arguments, and \emph{critical}
arguments chosen where the evaluation cancels (e.g., within a relative distance of $2^{-\PB}$ to $2^{-8}$
of a multiple of $\pi/2$ for the trigonometric functions, near $\pm1$ for \code{asin} and \code{acos},
near $1$ for \code{log}, and $|z|$ near 1 for the complex logarithm).
The test suite distributed with the library, on which the statement ``0~\ulp'' of the previous version of this paper
was based, used 3,000 to 5,000 arguments of order one per function, did not test division or the roots,
and did not contain critical arguments.

\paragraph{Run-time-precision layer}
The test program generates the test cases on the host with the system GMP and passes the operands to the
GPU as exact triples (sign, exponent, limbs) through the MPFR custom interface. Each GPU thread executes
its cases with \code{cu\_mpfr}/\code{cu\_mpc} and exports, for every case, the kind of the result (NaN,
$\pm\infty$, $\pm0$, or a regular number), its sign, its exponent, and all limbs, together with the ternary
return value and the MPFR flags raised by the operation. The same cases are evaluated on the CPU with the
system MPFR~4.2.2 on GMP~6.3.0 and a separately built MPC~1.4.1 (the versions that were ported), and the two
sides are compared without any tolerance and without passing through binary64. The cases cover the
following.
\begin{itemize}
  \item \textbf{Operations.} $+,-,\times,\div,\sqrt{\ }$ and the 12 real functions of
        Table~\ref{tab:elem_real} for \code{cu\_mpfr}; $+,-,\times,\div$ and the 12 complex functions of
        Table~\ref{tab:elem_cplx} for \code{cu\_mpc}.
  \item \textbf{Precisions.} 21 values from 2 to 2048 bits for the real basic operations, including values
        just below, at, and just above multiples of 64 (63, 64, 65, 127, 128, 129, 255, 256, 257); 24 to 2048
        bits for the real functions; 2 to 2048 bits for the complex basic operations and 53 to 1024 bits for
        the complex functions.
  \item \textbf{Rounding.} \code{MPFR\_RNDN}, \code{RNDZ}, \code{RNDU}, \code{RNDD}, and \code{RNDA}, chosen
        independently for each case and, for complex operations, for each of the two parts.
        \code{MPFR\_RNDF} is tested separately for the real basic operations: because its result is not
        unique, a GPU result is accepted if it equals the \code{RNDD} or the \code{RNDU} result of the
        reference, and the ternary value and flags are not compared.
  \item \textbf{Inputs.} (i)~Random significands. (ii)~Operands whose precisions differ from each other and
        from the precision of the result, in both directions. (iii)~Significands consisting of long runs of
        zeros and ones (GMP's \code{mpz\_rrandomb}). (iv)~Constructed results that are exactly representable
        or exact midpoints, and their neighbors: for $a\pm b$, $b$ equal to half a unit in the last place of
        $a$ and the two numbers adjacent to it; for $a\times b$, operand precisions that sum to $\PB+1$; for
        $u\div v$, $u=v\cdot c$ computed exactly with $c$ of \PB{} or $\PB+1$ bits; for $\sqrt{u}$, $u=c^2$
        likewise. (v)~Special values ($\pm0$, $\pm\infty$, NaN, $\pm1$, powers of two) in every operand
        position. (vi)~Exponent range: a range set by the thread for the case ($[-16,16]$, $[-64,64]$,
        $[-1000,1000]$; for complex operations $[-10^3,10^3]$ and $[-10^5,10^5]$) with operands chosen so
        that the result falls near its ends, and the default range with operand exponents within a few units
        of $\pm(2^{30}-1)$, so that overflow and underflow occur. (vii)~For the functions, arguments of
        order one, tiny arguments (exponents down to $-2\PB-10$), and large arguments (up to $2^{31}$).
\end{itemize}
Hard-to-round arguments of the elementary functions cannot be constructed in general; they are covered
only to the extent that the random and structured inputs happen to contain them.

\paragraph{Concurrent execution}
Every launch distributes its cases over 8192 threads (32 blocks of 256). Because the operation, the
precision, the rounding mode, the input class, and the exponent range are drawn independently for each
case, the threads of one warp execute different operations at different precisions with different
exponent ranges at the same time; in the \emph{mixed} runs, real and complex operations and basic
operations and functions are drawn into the same kernel. After each kernel, the state block of every thread
(Section~\ref{sec:arena}) is read back and checked: its flags must be those of the last case executed by
that thread, and its exponent range must again be the default one.

To confirm that this test detects the race described in Section~\ref{sec:arena}, we ran the mixed cases
also with the state blocks not installed, that is, with one state shared by all threads as in the
previous version, using only cases in the default exponent range.

\subsection{Results}

Table~\ref{tab:verify} summarizes the verification on GB10. The same tests were run on H100 (\code{sm\_90})
with the same outcome: no mismatch in the 27,000,000 basic operations of the fixed-precision layer, a summary
of the errors of division and of the functions that is identical to Table~\ref{tab:fixed_ulp} in every
entry, and no discrepancy in the 3,300,000 cases of the run-time-precision layer under the acceptance
criterion of each rounding mode described below, with all 431,360 thread state blocks as expected. (The randomly drawn cases are distributed slightly
differently over the rows of the table in the two runs; the totals are the same.)

\begin{table*}[t]
\centering
\footnotesize
\caption{Scope and results of the numerical verification against the system MPFR~4.2.2/MPC~1.4.1 (GB10).
For the first two and the last five rows, a mismatch is any difference in the kind, sign, exponent, or
significand of a result; for the \code{cu\_mpfr}/\code{cu\_mpc} rows with the rounding modes N, Z, U, D,
and A, the ternary value and the flags were compared as well. For \code{RNDF}, only membership in the set of
admissible results (the \code{RNDD} and the \code{RNDU} result of the reference) was checked; ternary
values and flags were not compared. The 1024-bit row was run on GB10 only and is not included in the row
above it. Input classes (i)--(vii) are defined in the text. The counts in parentheses are the
additional cases of the mixed (concurrent) runs.}
\label{tab:verify}
\setlength{\tabcolsep}{3pt}
\begin{tabular}{@{}lp{0.17\textwidth}p{0.10\textwidth}p{0.10\textwidth}p{0.13\textwidth}rp{0.10\textwidth}@{}}
\toprule
layer & operations & bits & rounding & inputs & cases & result \\
\midrule
\freal    & $+,-,\times$ & 32--2048 (15 values) & \rndn & random, corner cases & 18,000,000 & 0 mismatches \\
\fcomplex & $+,-,\times$ & 32--2048 (15 values) & \code{RNDNN} & random, corner cases & 9,000,000 & 0 mismatches \\
\freal/\fcomplex & $\div$, $\sqrt{\ }$, cube root, 14 real and 8 complex functions & 64, 128, 256, 512 & (none) &
  order one, tiny, large, critical & 41,256,099 & Table~\ref{tab:fixed_ulp} \\
\freal/\fcomplex & real and complex $\div$ & 1024 & (none) & order one, tiny, large, critical & 800,000 & 0 mismatches (rounded reference) \\
\midrule
\code{cu\_mpfr} & $+,-,\times,\div,\sqrt{\ }$ & 2--2048 (21 values) & N, Z, U, D, A & (i)--(vi) & 1,749,878 (+218,642) & 0 mismatches \\
\code{cu\_mpfr} & $+,-,\times,\div,\sqrt{\ }$ & 2--2048 (21 values) & \code{RNDF} & (i)--(vi) & 250,122 (+31,458) & 0 inadmissible results \\
\code{cu\_mpfr} & 12 real functions & 24--2048 & N, Z, U, D, A & (i)--(iii), (v)--(vii) & 400,000 (+99,996) & 0 mismatches \\
\code{cu\_mpc}  & $+,-,\times,\div$ & 2--2048 & 25 pairs of modes & (i)--(iii), (v), (vi) & 200,000 (+50,108) & 0 mismatches \\
\code{cu\_mpc}  & 12 complex functions & 53--1024 & 25 pairs of modes & (i)--(iii), (v)--(vii) & 200,000 (+99,796) & 0 mismatches \\
\bottomrule
\end{tabular}
\end{table*}

\paragraph{Run-time-precision layer}
No discrepancy was found in any of the 3,300,000 cases (2,800,000 in the runs organized by row and 500,000
in the mixed runs), under the acceptance criterion of each rounding mode. In the 3,018,420 cases with the
deterministic rounding modes (N, Z, U, D, A), the value, the ternary value, and the flags returned by
\code{cu\_mpfr}/\code{cu\_mpc} on the GPU were identical to those of the system libraries. In the 281,580
cases with \code{MPFR\_RNDF}, every value was equal to the \code{RNDD} or the \code{RNDU} result of the
reference; this is the acceptance criterion of faithful rounding, not bit-identity with the reference, and
the ternary value and the flags were not compared for these cases. All 431,360 thread state blocks inspected after the kernels held the
default exponent range and the expected flags, namely those that the reference raised in the last case
executed by the thread; for a thread whose last case used \code{MPFR\_RNDF}, only the exponent range was
checked. For the operations, precisions, rounding modes, and input
classes of Table~\ref{tab:verify}, and for the deterministic rounding modes, the run-time-precision layer
thus reproduces upstream MPFR/MPC bit for bit, including the exception flags and the behavior at the limits
of the exponent range. This is a
statement about the tested range; it is not a proof for all inputs.

Agreement with upstream includes agreement where upstream itself deviates from correct rounding, and the
two must be distinguished. While constructing the midpoint cases, we observed that \code{mpfr\_div} of the
unpatched MPFR~4.2.2 returns the odd neighbor for some exact midpoints when the precision of the quotient
is $63 \bmod 64$ bits and the divisor is sufficiently longer than the quotient: 4.5\% to 5.2\% of
20,000 random midpoints for 63-bit quotients with divisors of 113 bits or more, for 127- and 191-bit
quotients with divisors of 200 bits or more, and for 255- and 319-bit quotients with 1000-bit divisors,
and none for the other precisions that we tried (31, 62, 64, 65, 126, and 128 bits). The
observation is made with a program that uses only GMP and MPFR and checks the quotient with exact integer
arithmetic; one instance, the conditions, and the cause are given in Appendix~\ref{sec:appendix_tie}. We
reported the observation to the MPFR developers. The ported code
returns the same values as the system library in these cases, as expected of the same code, so that they
count as agreements in Table~\ref{tab:verify}. The claim of this section is therefore agreement with
MPFR~4.2.2 and MPC~1.4.1, not mathematically correct rounding in every case.

\paragraph{Control experiment with a shared state}
Table~\ref{tab:shared} shows the control runs in which all threads share one state, as in the version
used for the previous version of this paper. With one thread, all 50,000 cases agree with the reference. With two
threads, 12\% of the cases return wrong flags, and with 32 threads, 68\%; with 8192 threads, the kernel
aborts with a device error. The values and ternary values were not affected in the runs that completed,
but a kernel that aborts returns no values at all. The test therefore detects the race, and the absence
of any mismatch with per-thread state blocks is evidence that the race has been removed and not merely
that it went unobserved. On H100 the same control runs gave wrong flags in 7,054 (two threads) and 30,543
(32 threads) of the 50,000 cases and, with 32 threads, one wrong value; with 8192 threads the kernel aborted
as on GB10.

\begin{table}[htb]
\centering
\caption{Control experiment: mixed test cases (default exponent range) executed with the MPFR state shared
by all GPU threads (GB10).}
\label{tab:shared}
\footnotesize\setlength{\tabcolsep}{4pt}
\begin{tabular}{rrrrr}
\toprule
threads & cases & value & ternary & flags \\
\midrule
1    & 50,000 & 0 & 0 & 0 \\
2    & 50,000 & 0 & 0 & 6,039 \\
32   & 50,000 & 0 & 0 & 34,026 \\
8192 & \multicolumn{4}{c}{kernel aborted (device error)} \\
\bottomrule
\end{tabular}
\end{table}

\paragraph{Fixed-precision layer: basic operations}
All 27,000,000 results of addition, subtraction, and multiplication of \freal{} and \fcomplex{} were
identical to the MPFR/MPC round-to-nearest results.

\paragraph{Fixed-precision layer: division and functions}
Table~\ref{tab:fixed_ulp} gives the largest observed errors. The error of a real result is measured in
\ulp{} of the \PB-bit result; for a complex result it is measured separately for the real
and the imaginary part, each in its own \ulp, and the larger of the two is reported, so that a part that
nearly vanishes is held to its own unit. Independently of the printed error, we counted the results that
differ from the rounded reference. Three groups must be distinguished.
\begin{itemize}
  \item \emph{Group 1}: division, square root, cube root, \code{exp}, \code{expm1}, \code{log1p},
        \code{atan}, \code{sinh}, \code{cosh}, \code{atanh}, \code{pow}, complex division, and complex
        square root. All 20,456,099 results, in every class and at every precision, agreed with the
        rounded reference. For the real and the complex division, which the root-finding programs
        use at up to 1024 bits, we repeated the test at 1024 bits on GB10: all 800,000 results (100,000
        per class) agreed with the rounded reference as well. These evaluations involve no cancellation that the 128
        guard bits do not cover; the observation is consistent with an error of $\tfrac12$~\ulp{} plus a
        term of the order of $2^{-120}$~\ulp, but we have not proved such a bound, and exact ties and
        near-ties of division are covered only by the constructed cases of the critical class.
  \item \code{log}: 88 of the 1,600,000 results (all at 128 bits and for arguments near 1) differed from the
        rounded reference; their errors exceed 0.5~\ulp{} by less than $10^{-6}$~\ulp.
  \item \emph{Functions whose evaluation cancels at particular arguments.} For these, the results agreed with the
        rounded reference for all arguments of order one, tiny arguments, and large arguments (up to
        $2^{60}$ for the trigonometric functions), with one exception, the complex tangent, but not in
        the critical class, where only about 128 bits of the result are significant regardless of \PB. The
        failures that we observed are the following.
        (a)~\code{sin}, \code{cos}, \code{tan} and the complex \code{exp}, \code{sin}, \code{cos},
        \code{tan}, \code{sinh}, \code{cosh}, which use them, for arguments within a relative distance of
        $2^{-\PB}$ to $2^{-8}$ of a multiple of $\pi/2$, because the argument is reduced with a
        $(\PB+128)$-bit value of $\pi$: the error is at most 0.5~\ulp{} at 64 bits but reaches
        $2\times10^{3}$~\ulp{} at 128 bits, $3\times10^{41}$~\ulp{} at 256 bits, and
        $9\times10^{118}$~\ulp{} at 512 bits.
        (b)~\code{acos} near $\pm1$ from 128 bits (4 results differing from the rounded reference at 128 bits, errors up
        to $2\times10^{18}$ and $8\times10^{56}$~\ulp{} at 256 and 512 bits) and \code{asin} near $\pm1$
        at 512 bits ($4\times10^{8}$~\ulp).
        (c)~The complex logarithm for $|z|$ near 1, from 128 bits (up to $3\times10^{120}$~\ulp{} at 512
        bits).
        (d)~The complex tangent also for ordinary arguments with $4\le|\mathrm{Im}\,z|<64$ and a large real part, at every
        precision: 15\% of the results of the class of large arguments are wrong, with errors up to
        $10^{17}$~\ulp.
\end{itemize}
The tests locate these failures but do not delimit them: we have not determined how close to a zero, a
pole, or a branch point an argument may come before the error exceeds a given bound, and we give no safe
region. The functions of the third group should therefore be used in the fixed-precision layer only when
the arguments are known to stay away from the points listed above; the run-time-precision layer has no such
restriction. None of the experiments in this paper uses the fixed-precision elementary functions: the
root-finding programs of Section~\ref{sec:rootfinding} use only $+,-,\times$, and complex division.

\begin{table}[htb]
\centering
\caption{Fixed-precision layer: largest observed error [\ulp] against MPFR/MPC over 100,000 inputs per
class and precision (GB10). ``ordinary'': arguments of order one, tiny, and large; ``critical'': arguments
where the evaluation cancels. Group~1: $\div$, $\sqrt{\ }$, cube root, \code{exp}, \code{expm1},
\code{log1p}, \code{atan}, \code{sinh}, \code{cosh}, \code{atanh}, \code{pow}, complex $\div$
and $\sqrt{\ }$. For complex functions, the larger of the errors of the real and the imaginary part.
$^{\ast}$: some results differ from the rounded reference although the largest error is printed as 0.5.}
\label{tab:fixed_ulp}
\footnotesize\setlength{\tabcolsep}{3pt}
\begin{tabular}{@{}p{0.34\columnwidth}lrrrr@{}}
\toprule
functions & arguments & 64 & 128 & 256 & 512 \\
\midrule
group 1 & ordinary & 0.5 & 0.5 & 0.5 & 0.5 \\
 & critical & 0.5 & 0.5 & 0.5 & 0.5 \\
\code{log} & ordinary & 0.5 & 0.5 & 0.5 & 0.5 \\
 & critical & 0.5 & 0.5$^{\ast}$ & 0.5 & 0.5 \\
\code{sin}, \code{cos}, \code{tan} & ordinary & 0.5 & 0.5 & 0.5 & 0.5 \\
 & critical & 0.5 & 2\mbox{e}3 & 3\mbox{e}41 & 7\mbox{e}118 \\
\code{asin} & ordinary & 0.5 & 0.5 & 0.5 & 0.5 \\
 & critical & 0.5 & 0.5 & 0.5 & 4\mbox{e}8 \\
\code{acos} & ordinary & 0.5 & 0.5 & 0.5 & 0.5 \\
 & critical & 0.5 & 0.5$^{\ast}$ & 2\mbox{e}18 & 8\mbox{e}56 \\
complex \code{log} & ordinary & 0.5 & 0.5 & 0.5 & 0.5 \\
 & critical & 0.5 & 7\mbox{e}4 & 4\mbox{e}43 & 3\mbox{e}120 \\
complex \code{exp}, \code{sin}, \code{cos}, \code{sinh}, \code{cosh} & ordinary & 0.5 & 0.5 & 0.5 & 0.5 \\
 & critical & 0.5 & 2\mbox{e}3 & 3\mbox{e}41 & 9\mbox{e}118 \\
complex \code{tan} & ordinary & 1\mbox{e}17 & 1\mbox{e}17 & 8\mbox{e}16 & 8\mbox{e}16 \\
 & critical & 0.5 & 2\mbox{e}3 & 3\mbox{e}41 & 9\mbox{e}118 \\
\bottomrule
\end{tabular}
\end{table}

\subsection{Scope of the guarantees}

The statements made in the remainder of this paper are limited as follows.
\begin{itemize}
  \item Results of \freal/\fcomplex{} addition, subtraction, and multiplication are bit-identical to MPFR/MPC
        round-to-nearest within the tested range of Table~\ref{tab:verify}, for finite, nonzero values
        whose exponents stay in range; the type has no exceptional values.
  \item Results of \freal/\fcomplex{} division and functions are not claimed to be correctly rounded or
        faithful. Consequently, an algorithm that uses them (the root-finding iterations of
        Section~\ref{sec:rootfinding} use complex division) can follow a different trajectory from the same
        algorithm executed with MPFR/MPC.
  \item For \code{cu\_mpfr}/\code{cu\_mpc}, bit-identical agreement with upstream, including the ternary
        value and the flags, is claimed for the operations, precisions, rounding modes, and execution
        conditions of Table~\ref{tab:verify}, with per-thread state blocks installed, and for the rounding
        modes N, Z, U, D, and A. For \code{MPFR\_RNDF} only the admissibility of the value is claimed.
        Agreement with upstream is not the same as correct rounding: where upstream deviates from it
        (Appendix~\ref{sec:appendix_tie}), so does the port.
  \item When the order of operations differs between the CPU and GPU implementations of an algorithm (e.g.,
        a tree reduction on the GPU versus a sequential sum on the CPU), the results are not bit-identical
        even if every individual operation is.
\end{itemize}

\section{Performance of arithmetic kernels}
\label{sec:benchmarktest}

\subsection{Environment and measurement conditions}
\label{sec:env}

We used the following two machines, which we refer to by the names of their GPUs.
\begin{description}
  \item[GB10] NVIDIA DGX Spark~\cite{nvidia_dgx_spark}: Arm CPU with 10 Cortex-X925 and 10 Cortex-A725 cores,
        128\,GB LPDDR5X memory shared by the CPU and the GPU, NVIDIA GB10 GPU (\code{sm\_121}), CUDA~13.0,
        Ubuntu 24.04 LTS, GCC 13.3.0.
  \item[H100] Intel Xeon Gold 6526Y $\times 2$ (32 cores in total), 1\,TB RAM, NVIDIA H100 NVL
        (94\,GB HBM3, 132 SMs)~\cite{nvidia_h100} $\times 4$ (\code{sm\_90}), CUDA~13.0, Ubuntu 24.04 LTS, GCC 13.3.0.
        All measurements in this paper use one of the four GPUs.
\end{description}

\paragraph{Measurement}
GPU times were measured with \code{cudaEventRecord} after one warm-up launch, which excludes first-launch
compilation, context creation, and the setting of the device limits. Each reported GPU time is the mean
of 10 timed launches (20 in Section~\ref{sec:cump}) of the kernel alone; host--device transfer is excluded.
The sample standard deviation is reported or bounded in each subsection. All operands are random and use
the full significand; they are generated once on the host and passed to the device as exact (sign,
exponent, limbs) triples. Every GPU result in Sections~\ref{sec:elementary_function}
and~\ref{sec:blas} was compared bit by bit with a \emph{correctness reference}: the result of the system
MPFR/MPC executing the same operations in the same order as the GPU kernel (for the block-per-row
matrix--vector product, the same partial sums and the same tree). No difference was found. The correctness
reference is distinct from the \emph{timed CPU baselines} described below, which compute the same quantities
in the natural sequential order of a CPU implementation.

The benchmark programs of the previous version of this paper differed in three respects that affect the numbers:
their operands were binary64 values converted inside the timed kernel, each time was a single launch, and
their ``CPU'' time was the host path of mpc\_cuda compiled without host optimization (nvcc passes no
optimization option to the host compiler unless one is given). The numbers of this paper therefore replace,
and are not comparable with, those of the previous version of this paper.

\paragraph{CPU baselines}
Three timed CPU baselines are used in Sections~\ref{sec:elementary_function} and~\ref{sec:blas}, each
computing the same quantities from the same operands as the GPU kernel; their order of summation is the
sequential one, which differs from that of the GPU only for the block-per-row matrix--vector product.
\begin{itemize}
  \item \textbf{sys(1), sys(all)}: the system MPFR~4.2.2/MPC~1.4.1 on GMP~6.3.0, which use GMP's assembly
        kernels, with one thread and with one OpenMP thread per core over the independent work items
        (20 threads on GB10, 32 on H100; \code{OMP\_PROC\_BIND=close}, \code{OMP\_PLACES=cores}). The
        speedups quoted in the abstract and the conclusion are relative to sys(all), the strongest of the
        baselines.
  \item \textbf{host(1)}: the host path of mpc\_cuda, that is, the same converted mini-GMP-based sources
        compiled for the CPU with \code{-O2}, single-threaded. The ratio of host(1) to the GPU time shows
        the effect of executing the same code on the GPU; it is not a speedup over an efficient CPU
        implementation, because the mini-GMP-based code is several times slower than the system libraries.
\end{itemize}
The conditions of each table are collected in Appendix~\ref{sec:appendix_repro}.

\subsection{Comparison with existing GPU libraries: AXPY}
\label{sec:cump}

We compare both layers of mpc\_cuda with CUMP~\cite{cump} and CAMPARY~\cite{campary_paper2016}
(Section~\ref{sec:related}). Because CUMP implements only addition, subtraction, and multiplication, the
comparison uses the real AXPY $\mathbf{y}\leftarrow a\mathbf{x}+\mathbf{y}$, which all four implementations
can execute under the same conditions, with $N=10^5$ and precisions from 128 to 16384 bits.

\paragraph{Conditions}
All four programs draw the same sequence of random \PB-bit values $a, x_i, y_i\in[0,1)$ from GMP's generator
with a common seed, so that the inputs use the full significand and are identical across libraries. Each
library's result is returned to the host at full precision and compared with a reference
$a x_i+y_i$ computed with $\PB+64$ bits. All kernels write the result to a separate output vector, so that
every launch does the same work. Each time in Table~\ref{tab:cump} is the mean of 20 timed launches after
one warm-up launch; on GB10 the sample standard deviation was at most 8.2\% of the mean (CAMPARY, 256 bits)
and at most 6.4\% for the other entries. Each library is launched in the configuration of its own example
or test programs: CUMP with 3125 blocks of 32 threads, CAMPARY and \freal{} with 391 blocks of 256 threads
(one element per thread), and \code{cu\_mpfr} with 8192 threads (128 blocks of 64) that stride over the
elements, each with a 256\,KB arena slab.

A nominal precision of \PB{} bits does not mean the same working precision in every library
(Table~\ref{tab:prec_map}). \code{cu\_mpfr} and \freal{} carry exactly \PB{} bits and round to nearest.
CAMPARY carries $K=\lceil \PB/53\rceil$ binary64 components. CUMP follows GMP's \code{mpf}: for a requested
precision of \PB{} bits it sets the precision to $\PB/64+1$ limbs of 64 bits and stores and computes
$\PB/64+2$ limbs, which we confirmed by reading the sizes of the result objects. The comparison is
therefore between implementations that provide \emph{at least} \PB{} bits, and the accuracy actually
obtained is reported separately in Table~\ref{tab:acc}.

\begin{table}[htb]
\centering
\caption{Working precision behind each nominal precision in the AXPY comparison.}
\label{tab:prec_map}
\footnotesize\setlength{\tabcolsep}{3pt}
\begin{tabular}{rrrrr}
\toprule
nominal & \multicolumn{2}{c}{CAMPARY} & CUMP & \code{cu\_mpfr}, \freal \\
\cmidrule(lr){2-3}
bits & $K$ & effective bits & limbs (bits) & limbs (bits) \\
\midrule
128 & 3 & 159 & 4 (256) & 2 (128) \\
256 & 5 & 265 & 6 (384) & 4 (256) \\
512 & 10 & 530 & 10 (640) & 8 (512) \\
1024 & 20 & 1060 & 18 (1152) & 16 (1024) \\
2048 & 39 & 2067 & 34 (2176) & 32 (2048) \\
4096 & 78 & 4134 & 66 (4224) & 64 (4096) \\
8192 & 155 & 8215 & 130 (8320) & 128 (8192) \\
16384 & (310) & --- & 258 (16512) & 256 (16384) \\
\bottomrule
\end{tabular}
\end{table}

\begin{table*}[t]
\footnotesize
\centering
\caption{GPU time of multiple-precision AXPY [ms] ($N=10^5$, mean of 20 launches). \code{cu\_mpfr} is the
run-time-precision layer and \freal{} the fixed-precision layer of mpc\_cuda. The fastest entry in each row
and machine is in bold. $-$: the kernel could not be compiled.}
\label{tab:cump}
\begin{tabular}{rrrrr@{\hskip 3em}rrrr}
\toprule
 & \multicolumn{4}{c}{GB10} & \multicolumn{4}{c}{H100} \\
\cmidrule(r){2-5}\cmidrule(l){6-9}
bits & CUMP & CAMPARY & \code{cu\_mpfr} & \code{cu\_freal} & CUMP & CAMPARY & \code{cu\_mpfr} & \code{cu\_freal} \\
\midrule
128 & 0.044 & 0.113 & 0.309 & \textbf{0.011} & 0.013 & 0.018 & 0.157 & \textbf{0.008} \\
256 & 0.096 & 0.277 & 0.497 & \textbf{0.019} & 0.016 & 0.033 & 0.238 & \textbf{0.010} \\
512 & 0.227 & 0.848 & 0.823 & \textbf{0.035} & 0.030 & 0.114 & 0.321 & \textbf{0.016} \\
1024 & 0.475 & 3.706 & 1.770 & \textbf{0.193} & 0.064 & 0.449 & 0.559 & \textbf{0.041} \\
2048 & 1.210 & 17.02 & 10.78 & \textbf{0.457} & 0.201 & 2.568 & 2.339 & \textbf{0.090} \\
4096 & \textbf{5.058} & 53.93 & 56.92 & 7.871 & \textbf{0.829} & 9.304 & 9.631 & 1.222 \\
8192 & \textbf{23.55} & 244.4 & 191.7 & 195.9 & \textbf{3.385} & 77.60 & 30.34 & 5.446 \\
16384 & \textbf{130.5} & $-$ & 752.4 & 336.2 & \textbf{13.66} & $-$ & 170.1 & 70.55 \\
\bottomrule
\end{tabular}
\end{table*}

\begin{table}[htb]
\centering
\caption{Maximum relative error of the AXPY results against a reference computed with $\PB+64$ bits
($N=10^5$, GB10).}
\label{tab:acc}
\footnotesize\setlength{\tabcolsep}{4pt}
\begin{tabular}{rllll}
\toprule
bits & CUMP & CAMPARY & \code{cu\_mpfr} & \code{cu\_freal} \\
\midrule
128 & $2.2\mbox{e}-55$ & $1.7\mbox{e}-48$ & $4.3\mbox{e}-39$ & $4.3\mbox{e}-39$ \\
256 & $1.2\mbox{e}-94$ & $1.9\mbox{e}-80$ & $1.3\mbox{e}-77$ & $1.3\mbox{e}-77$ \\
512 & $2.6\mbox{e}-170$ & $1.9\mbox{e}-161$ & $1.1\mbox{e}-154$ & $1.1\mbox{e}-154$ \\
1024 & $2.9\mbox{e}-325$ & $1.3\mbox{e}-320$ & $7.8\mbox{e}-309$ & $7.8\mbox{e}-309$ \\
2048 & $5.7\mbox{e}-634$ & $4.1\mbox{e}-321$ & $4.6\mbox{e}-617$ & $4.6\mbox{e}-617$ \\
4096 & $1.1\mbox{e}-1250$ & $4.2\mbox{e}-321$ & $1.4\mbox{e}-1233$ & $1.4\mbox{e}-1233$ \\
8192 & $1.4\mbox{e}-2483$ & $1.7\mbox{e}-320$ & $1.4\mbox{e}-2466$ & $1.4\mbox{e}-2466$ \\
16384 & $8.8\mbox{e}-4950$ & $-$ & $1.2\mbox{e}-4932$ & $1.2\mbox{e}-4932$ \\
\bottomrule
\end{tabular}
\end{table}

\paragraph{Accuracy}
Table~\ref{tab:acc} shows the errors on GB10; those on H100 are identical in every entry. The errors of \code{cu\_mpfr} and \freal{} coincide and are of the
order of $2^{-\PB}$, as expected for two correctly rounded operations. CUMP and, up to 512 bits, CAMPARY are
more accurate than requested because they carry more bits (Table~\ref{tab:prec_map}); the 128-bit row in
particular compares a 128-bit result with results that carry 159 bits (CAMPARY) and 256 bits (CUMP). The
error of CAMPARY does not decrease below approximately $10^{-320}$: for values of order one, the low-order
components of an expansion longer than about 1000 bits fall below the smallest binary64 subnormal number.
Above 1024 bits, the CAMPARY entries of Table~\ref{tab:cump} therefore do not deliver the nominal precision
and are listed only for reference.

\paragraph{Time}
Table~\ref{tab:cump} leads to the following observations, which hold on both machines unless stated
otherwise.
\begin{itemize}
  \item \textbf{\freal{} versus CUMP.} \freal{} is faster than CUMP up to 2048 bits, by factors of
        $2.5$--$6.5\times$ on GB10 and $1.6$--$2.2\times$ on H100. From 4096 bits CUMP is faster: by
        $1.6\times$, $8.3\times$, and $2.6\times$ at 4096, 8192, and 16384 bits on GB10, and by $1.5\times$,
        $1.6\times$, and $5.2\times$ on H100.
  \item \textbf{\freal{} versus CAMPARY.} \freal{} is faster than CAMPARY at every precision at which both
        run, by factors from $1.2\times$ (8192 bits) to $37\times$ (2048 bits) on GB10 and from $2.3\times$
        (128 bits) to $29\times$ (2048 bits) on H100. The 16384-bit CAMPARY kernel ($K=310$) could not be
        compiled because the template recursion depth of its renormalization exceeds the limit of nvcc.
  \item \textbf{\freal{} versus \code{cu\_mpfr}.} \freal{} is $28\times$ (GB10) and $21\times$ (H100)
        faster than the run-time-precision layer at 128 bits, and $9.2\times$ and $14\times$ faster at
        1024 bits. At 8192 bits the two take the same time on GB10 (196 and 192\,ms), whereas \freal{}
        remains $5.6\times$ faster on H100.
  \item \textbf{\code{cu\_mpfr} versus CUMP.} The run-time-precision layer is slower than CUMP at every
        precision, by factors of $3.6$--$11\times$ on GB10 and $8.7$--$15\times$ on H100.
  \item \textbf{GB10 versus H100.} Every entry is smaller on H100: by factors of $3.5$--$9.6$ for CUMP,
        $3.2$--$8.4$ for CAMPARY, $2.0$--$6.3$ for \code{cu\_mpfr}, and $1.5$--$6.4$ for \freal, with the
        exception of \freal{} at 8192 bits ($36\times$), which is discussed below.
\end{itemize}
These statements concern the three libraries measured here and one kernel. In particular, the advantage of
\freal{} at 128 and 256 bits is an advantage over CUMP, CAMPARY, and \code{cu\_mpfr}; it says nothing about
the libraries specialized for at most 212 bits~\cite{gqd,mx_real,cuda_fp128}, which we did not measure.

The \code{cu\_mpfr} column differs from the one of the previous version of this paper, which timed a kernel whose
operands were set from binary64 values inside the kernel. Its time was proportional to the precision
(0.96\,ms at 128 bits to 94\,ms at 16384 bits on GB10 when remeasured), because \code{mpfr\_mul} removes
low-order zero limbs from its operands before a long multiplication; with operands that use the full
significand, the same layer needs 752\,ms at 16384 bits. The old numbers therefore understated the cost of
the run-time-precision layer at high precision by up to a factor of 8 and overstated it at low precision,
where the conversion from binary64 and the initialization of four temporaries per element dominated.

\paragraph{Dependence on precision}
The three limb-based implementations use schoolbook multiplication, whose cost is quadratic in the number of
limbs, but their measured times scale differently. The time of CUMP increases by a factor of $11$ (GB10)
and $5.0$ (H100) from 128 to 1024 bits and reaches a factor of 4 or more per doubling only above 2048 bits. The
time of \code{cu\_mpfr} increases by a factor of 1.3 to 2.2 per doubling up to 1024 bits, by a factor of 6.1
(GB10) and 4.2 (H100) from 1024 to 2048 bits, and by 3.2 to 5.6 beyond. The step between 1024 and 2048 bits
coincides with the threshold
(\code{MPFR\_MUL\_THRESHOLD}, 20 limbs in the generic parameter set used by the port) above which
\code{mpfr\_mul} switches from its direct product to the code path based on a short product with temporary
storage; we did not measure the two paths separately.

The time of \freal{} increases much faster than the quadratic cost of multiplication explains: on GB10 by
a factor of 17 from 2048 to 4096 bits and by a factor of 25 from 4096 to 8192 bits, and on H100 by factors
of 14, 4.5, and 13 for the three doublings from 2048 to 16384 bits. Two measurements bear on the cause.
First, Table~\ref{tab:regs} lists the resources that the compiler reports for the kernel. Up to 1024 bits
no register spill is reported, although the kernel has a local frame of 48 to 416 bytes; at 2048 bits the kernel uses 254 (\code{sm\_121}) or all 255 (\code{sm\_90}) of the
registers available to a thread and the first spills appear; from 4096 bits the kernel is at the register
limit on both architectures and the amount of spill code grows by one to two orders of magnitude. We did not
inspect the generated machine code and therefore do not claim that all limbs are held in registers up to
1024 bits; what the compiler output shows is that register pressure causes no spills up to that precision
and a rapidly growing amount of spill code beyond it. Second, the device code of
\code{cu\_fmul} uses the truncated product of Section~\ref{sec:fixed} only up to 32 limbs and the full
$2N$-limb product above, a threshold that had been chosen on H100 with a different kernel. When the AXPY
kernel is recompiled with the truncated product enabled up to 128 limbs
(\code{-DCU\_FP\_DEV\_MULHIGH\_MAX\_N=128}), the times on GB10 become 2.99\,ms at 4096 bits and 22.6\,ms
at 8192 bits (mean of 10 launches, same errors as in Table~\ref{tab:acc}) instead of 7.87 and 196\,ms, which
is below the times of CUMP (5.06 and 23.6\,ms). On H100 the same change gives 0.92 and 4.74\,ms instead of
1.22 and 5.36\,ms, a gain of 25\% and 11\%, and CUMP remains faster (0.83 and 3.39\,ms). The largest part of
the jump on GB10 is thus a property of the code path selected by default, and the threshold of the released
header, which we kept for the tables, is not the best choice for this kernel on either machine. With the
better path, a step of $6.5\times$ (GB10) and $10\times$ (H100) remains between 2048 and 4096 bits, where
the kernel reaches the register limit; we attribute it to the spills of Table~\ref{tab:regs} but could not
confirm this with a profile (Section~\ref{sec:analysis}).
The number of launched threads also matters at these sizes, in opposite directions on the two machines: on
GB10, with 2048 threads striding over the elements instead of one thread per element, the default 8192-bit
kernel takes 96\,ms instead of 196\,ms, whereas on H100 it takes 36\,ms instead of 5.4\,ms, and no launch
configuration that we tried was faster there than one thread per element.

\begin{table}[htb]
\centering
\caption{Resources of the \freal{} AXPY kernel reported by the compiler (\code{ptxas -v}): registers per
thread, size of the per-thread local frame, and total size of the spill stores and loads.}
\label{tab:regs}
\footnotesize\setlength{\tabcolsep}{3pt}
\begin{tabular}{rrrrrrr}
\toprule
 & \multicolumn{3}{c}{\code{sm\_121} (GB10)} & \multicolumn{3}{c}{\code{sm\_90} (H100)} \\
\cmidrule(lr){2-4}\cmidrule(lr){5-7}
bits & reg. & frame [B] & spill [B] & reg. & frame [B] & spill [B] \\
\midrule
128 & 34 & 48 & 0 & 38 & 48 & 0 \\
256 & 54 & 80 & 0 & 54 & 80 & 0 \\
512 & 86 & 144 & 0 & 88 & 224 & 0 \\
1024 & 154 & 416 & 0 & 158 & 416 & 0 \\
2048 & 254 & 832 & 48 & 255 & 944 & 376 \\
4096 & 255 & 2368 & 6692 & 255 & 3184 & 8984 \\
8192 & 255 & 9664 & 71832 & 255 & 8880 & 116076 \\
16384 & 255 & 19904 & 41972 & 255 & 16512 & 230256 \\
\bottomrule
\end{tabular}
\end{table}

In summary, with respect to AXPY on GB10, the fixed-precision layer is the fastest of the four
implementations up to 2048 bits; CUMP is the fastest from 4096 bits but provides neither rounding guarantees
nor division; and the run-time-precision layer is slower than CUMP at every precision. The value of the
run-time-precision layer lies in its functionality (Table~\ref{tab:libs}), not in the speed of its basic
operations.

\subsection{MPRES-BLAS}
\label{sec:mpres}

MPRES-BLAS~\cite{mpres_blas_jpdc,mpres_blas} could be built on both architectures with CUDA~13.0 (one
compiler option was needed, Appendix~\ref{sec:appendix_repro}), and we measured its AXPY, GEMV, and GEMM on
the same GPU. It cannot, however, be put into Table~\ref{tab:cump}, for three reasons.
\begin{itemize}
  \item \textbf{Precision.} The precision is a property of the set of moduli selected at compile time.
        The predefined sets give 120, 240, 360, 480, 720, 960, 1200, 1440, 1680, 1920, 2160, and
        2400 bits; none of them equals a precision of Table~\ref{tab:cump}, and none exceeds 2400 bits.
  \item \textbf{Interface.} The operations are exposed as BLAS routines that launch several kernels from
        the host, with the launch configuration as an argument; we used the configurations of the
        library's own performance tests and did not tune them.
  \item \textbf{Semantics.} The significand is rounded by a scaling step based on an interval evaluation
        of the residue representation, which is not correct rounding; division and square root are
        computed on the host.
\end{itemize}
We therefore report MPRES-BLAS separately, with operands that carry exactly the precision of each moduli
set and are otherwise generated as in Section~\ref{sec:cump}, next to the mpc\_cuda results at the nearest
power of two (Table~\ref{tab:mpres}). The errors of the MPRES-BLAS results against MPFR were of the order of
$2^{-\PB}$ for its precision \PB{} in every completed run. Some runs aborted with an illegal memory access reported by the CUDA runtime: on GB10 with the two largest
sets (2160 and 2400 bits), and on H100 for AXPY and the matrix--vector product from the 960-bit set upward.
The failed entries of the table are runs that did not complete in three attempts. The failures are
intermittent (when the failed H100 cases were repeated with a single timed launch per process, one of them
completed in 5 of 19 attempts and the others in none of 20) and occurred equally with a device stack of
64\,KB instead of the default; we did not identify their cause.

\begin{table*}[t]
\centering
\caption{MPRES-BLAS 1.7.0 and mpc\_cuda on the same GPU: AXPY ($N=10^5$), matrix--vector product
($N=1024$), and matrix--matrix product ($N=96$), times in ms (mean of 10 launches; 20 for the mpc\_cuda
AXPY). The precisions of the two libraries do not coincide. For \code{cu\_mpfr}, the matrix--vector product
is the block-per-row kernel of Table~\ref{tab:linalg}. --: not measured; failed: see the text.}
\label{tab:mpres}
\footnotesize\setlength{\tabcolsep}{5pt}
\begin{tabular}{lr rrr @{\hskip 2em} rrr}
\toprule
 & & \multicolumn{3}{c}{GB10} & \multicolumn{3}{c}{H100} \\
\cmidrule(lr){3-5}\cmidrule(lr){6-8}
library & bits & AXPY & mat--vec & mat--mat & AXPY & mat--vec & mat--mat \\
\midrule
MPRES-BLAS & 120 & 22.0 & 181 & 178 & 16.1 & 209 & 226 \\
MPRES-BLAS & 240 & 22.0 & 190 & 198 & 29.7 & 217 & 242 \\
MPRES-BLAS & 480 & 33.3 & 280 & 251 & 37.1 & 262 & 318 \\
MPRES-BLAS & 960 & 64.1 & 566 & 434 & failed & failed & 736 \\
MPRES-BLAS & 1920 & 429 & 3357 & 1718 & failed & failed & 2138 \\
MPRES-BLAS & 2400 & failed & failed & 2662 & failed & failed & 3580 \\
\midrule
\code{cu\_mpfr} & 128 & 0.309 & -- & -- & 0.157 & -- & -- \\
\code{cu\_mpfr} & 256 & 0.497 & 4.09 & 3.18 & 0.238 & 1.00 & 1.55 \\
\code{cu\_mpfr} & 512 & 0.823 & -- & -- & 0.321 & -- & -- \\
\code{cu\_mpfr} & 1024 & 1.77 & 13.3 & 10.4 & 0.559 & 3.09 & 4.21 \\
\code{cu\_mpfr} & 2048 & 10.8 & -- & -- & 2.34 & -- & -- \\
\code{cu\_freal} & 128 & 0.011 & -- & -- & 0.008 & -- & -- \\
\code{cu\_freal} & 256 & 0.019 & -- & -- & 0.010 & -- & -- \\
\code{cu\_freal} & 512 & 0.035 & -- & -- & 0.016 & -- & -- \\
\code{cu\_freal} & 1024 & 0.193 & -- & -- & 0.041 & -- & -- \\
\code{cu\_freal} & 2048 & 0.457 & -- & -- & 0.090 & -- & -- \\
\bottomrule
\end{tabular}
\end{table*}

On GB10, MPRES-BLAS in this configuration is one to three orders of magnitude slower than mpc\_cuda at the
nearest precision: its AXPY at 960 bits takes 64.1\,ms, against 1.77\,ms for \code{cu\_mpfr} and 0.19\,ms
for \freal{} at 1024 bits, and its matrix products at 960 bits take 566\,ms and 434\,ms, against 13.3 and
10.4\,ms for \code{cu\_mpfr} at 1024 bits. We refrain from generalizing this result: the times of
MPRES-BLAS depend strongly on whether its rounding step is triggered (with 128-bit operands on the 240-bit
set, where the product needs no rounding, the same AXPY took 1.14\,ms instead of 22.0\,ms), the library was
developed and tuned on other GPUs, and its launch configurations were not adapted to GB10. The comparison
shows that, on this machine and without tuning, the residue number system does not give MPRES-BLAS an
advantage over the limb-based implementations for these kernels; it does not show how the libraries compare
on the hardware for which MPRES-BLAS was designed. On H100 the comparison is
possible only where MPRES-BLAS completed, and the result is the same: its AXPY at 480 bits takes 37.1\,ms,
against 0.32\,ms for \code{cu\_mpfr} and 0.016\,ms for \freal{} at 512 bits, and its matrix--matrix product
at 960 bits takes 736\,ms, against 4.2\,ms for \code{cu\_mpfr} at 1024 bits. Most MPRES-BLAS entries are
not smaller on H100 than on GB10, whereas all mpc\_cuda entries are, which we take as a further sign
that the launch configuration used here is not adapted to these GPUs.

GARPREC could not be measured, for the reasons given in Section~\ref{sec:related}.

\subsection{Real and complex elementary functions}
\label{sec:elementary_function}

We measured the elementary functions of the run-time-precision layer for $N=4096$ arguments at 1024 and
256 bits (Tables~\ref{tab:elem_real} and~\ref{tab:elem_cplx}). The real arguments are uniformly distributed
in $[0.5,1.5)$; the complex arguments have real parts in $[0.5,1.5)$ and imaginary parts in $[0.3,0.8)$; all
use the full significand, and all functions round to nearest. The kernel is launched with 256 blocks of 32
threads, one argument per thread, so that 4096 threads perform work, each with an arena slab of 1\,MB (real)
or 2\,MB (complex). The standard deviation of the GPU times was at most 6.0\% of the mean on GB10 and
0.9\% on H100. On both machines, all $2\times 24\times 4096$ results agree bit for bit with the system
MPFR/MPC.

\begin{table*}[t]
\centering
\caption{Real elementary functions of the run-time-precision layer \code{cu\_mpfr} ($N=4096$, 1024 bits),
times in ms. host(1): host path of mpc\_cuda, one thread; sys(1), sys(all): system MPFR with one thread and
with all cores (Section~\ref{sec:env}). The last two columns of each machine are the ratios of the host(1)
and sys(all) times to the GPU time (${>}1$: the GPU is faster).}
\label{tab:elem_real}
\footnotesize\setlength{\tabcolsep}{4pt}
\begin{tabular}{l rrrrrr @{\hskip 2em} rrrrrr}
\toprule
& \multicolumn{6}{c}{GB10} & \multicolumn{6}{c}{H100}\\
\cmidrule(lr){2-7}\cmidrule(lr){8-13}
func. & GPU & host(1) & sys(1) & sys(all) & $\frac{\mathrm{host(1)}}{\mathrm{GPU}}$ & $\frac{\mathrm{sys(all)}}{\mathrm{GPU}}$
      & GPU & host(1) & sys(1) & sys(all) & $\frac{\mathrm{host(1)}}{\mathrm{GPU}}$ & $\frac{\mathrm{sys(all)}}{\mathrm{GPU}}$ \\
\midrule
sqrt & 2.45 & 60.8 & 1.98 & 1.58 & 24.9 & 0.65 & 0.834 & 64.0 & 2.20 & 0.155 & 76.7 & 0.19 \\
cbrt & 8.53 & 230 & 5.21 & 1.36 & 27.0 & 0.16 & 3.48 & 225 & 6.06 & 0.482 & 64.7 & 0.14 \\
exp & 17.7 & 522 & 41.6 & 3.93 & 29.4 & 0.22 & 11.9 & 505 & 38.9 & 6.11 & 42.4 & 0.51 \\
expm1 & 17.6 & 538 & 44.0 & 3.36 & 30.5 & 0.19 & 11.2 & 513 & 40.8 & 3.77 & 45.8 & 0.34 \\
log & 83.1 & 1835 & 51.1 & 4.38 & 22.1 & 0.05 & 39.5 & 1881 & 51.9 & 5.22 & 47.6 & 0.13 \\
log1p & 72.7 & 1836 & 51.4 & 3.95 & 25.3 & 0.05 & 26.1 & 1881 & 52.1 & 2.69 & 72.0 & 0.10 \\
sin & 8.02 & 203 & 39.7 & 3.15 & 25.3 & 0.39 & 3.29 & 202 & 34.2 & 1.53 & 61.4 & 0.46 \\
cos & 4.90 & 130 & 34.1 & 3.93 & 26.6 & 0.80 & 2.11 & 112 & 29.1 & 2.00 & 53.3 & 0.95 \\
tan & 8.66 & 212 & 43.5 & 4.35 & 24.4 & 0.50 & 3.38 & 153 & 27.1 & 3.39 & 45.3 & 1.00 \\
atan & 42.8 & 957 & 152 & 9.55 & 22.3 & 0.22 & 19.6 & 717 & 98.3 & 15.4 & 36.6 & 0.79 \\
sinh & 30.0 & 545 & 45.4 & 4.27 & 18.2 & 0.14 & 26.0 & 375 & 30.6 & 2.86 & 14.4 & 0.11 \\
cosh & 21.6 & 538 & 45.9 & 3.73 & 24.9 & 0.17 & 14.9 & 369 & 29.8 & 3.41 & 24.7 & 0.23 \\
\bottomrule
\end{tabular}
\end{table*}

\begin{table*}[t]
\centering
\caption{Complex elementary functions of the run-time-precision layer \code{cu\_mpc} ($N=4096$, 1024 bits),
times in ms. Columns as in Table~\ref{tab:elem_real}, with the system MPC as the baseline.}
\label{tab:elem_cplx}
\footnotesize\setlength{\tabcolsep}{4pt}
\begin{tabular}{l rrrrrr @{\hskip 2em} rrrrrr}
\toprule
& \multicolumn{6}{c}{GB10} & \multicolumn{6}{c}{H100}\\
\cmidrule(lr){2-7}\cmidrule(lr){8-13}
func. & GPU & host(1) & sys(1) & sys(all) & $\frac{\mathrm{host(1)}}{\mathrm{GPU}}$ & $\frac{\mathrm{sys(all)}}{\mathrm{GPU}}$
      & GPU & host(1) & sys(1) & sys(all) & $\frac{\mathrm{host(1)}}{\mathrm{GPU}}$ & $\frac{\mathrm{sys(all)}}{\mathrm{GPU}}$ \\
\midrule
sqr & 0.456 & 9.08 & 3.36 & 1.95 & 19.9 & 4.29 & 0.239 & 10.5 & 4.50 & 0.358 & 43.8 & 1.50 \\
sqrt & 6.03 & 135 & 8.84 & 2.06 & 22.4 & 0.34 & 2.61 & 108 & 13.0 & 1.22 & 41.2 & 0.47 \\
exp & 45.6 & 1287 & 157 & 13.9 & 28.2 & 0.30 & 27.3 & 913 & 101 & 15.8 & 33.5 & 0.58 \\
log & 123 & 2732 & 216 & 12.3 & 22.1 & 0.10 & 71.8 & 2037 & 144 & 9.07 & 28.4 & 0.13 \\
sin & 36.3 & 784 & 94.2 & 7.01 & 21.6 & 0.19 & 32.2 & 551 & 61.5 & 3.25 & 17.1 & 0.10 \\
cos & 36.3 & 784 & 94.2 & 6.43 & 21.6 & 0.18 & 26.7 & 551 & 61.7 & 4.88 & 20.6 & 0.18 \\
tan & 49.8 & 866 & 120 & 8.81 & 17.4 & 0.18 & 43.9 & 612 & 76.8 & 7.07 & 13.9 & 0.16 \\
sinh & 40.1 & 783 & 93.2 & 7.29 & 19.5 & 0.18 & 33.7 & 553 & 61.0 & 5.70 & 16.4 & 0.17 \\
cosh & 36.0 & 782 & 93.3 & 6.86 & 21.8 & 0.19 & 27.7 & 549 & 61.2 & 6.10 & 19.8 & 0.22 \\
asin & 165 & 3625 & 268 & 15.6 & 22.0 & 0.09 & 91.7 & 2711 & 185 & 17.7 & 29.6 & 0.19 \\
acos & 189 & 3859 & 249 & 14.7 & 20.4 & 0.08 & 101 & 2884 & 168 & 7.79 & 28.7 & 0.08 \\
atan & 264 & 5866 & 442 & 24.3 & 22.2 & 0.09 & 130 & 4383 & 298 & 13.9 & 33.6 & 0.11 \\
\bottomrule
\end{tabular}
\end{table*}

The results at 1024 bits are as follows.
\begin{itemize}
  \item \textbf{Same code, GPU versus one CPU core.} The GPU is 18--31$\times$ (real functions) and
        17--28$\times$ (complex functions) faster than the host path on GB10, and 14--77$\times$ and
        14--44$\times$ faster on H100. This is the effect of the execution environment on identical sources.
  \item \textbf{Against the system libraries with one thread.} For the real functions, the GPU is between
        $0.6\times$ (\code{cbrt}) and $7.0\times$ (\code{cos}) as fast as the system MPFR on GB10 and
        between $1.2\times$ (\code{sinh}) and $14\times$ (\code{cos}) on H100; for the complex functions,
        between $1.3\times$ (\code{acos}) and $7.4\times$ (\code{sqr}) as fast as the system MPC on GB10
        and between $1.7\times$ (\code{acos}) and $19\times$ (\code{sqr}) on H100.
  \item \textbf{Against the system libraries with all cores.} On GB10 (20 cores) the GPU is slower for every
        real function (ratios 0.05 for \code{log} to 0.80 for \code{cos}) and for every complex function
        except \code{sqr} ($4.3\times$); the other complex ratios are between 0.08 and 0.34. On H100
        (32 cores) no real function is faster on the GPU (ratios 0.10 for \code{log1p} to 1.00 for
        \code{tan}), and the complex ratios are between 0.08 and 0.58 except for \code{sqr} ($1.5\times$).
\end{itemize}
At 256 bits the picture is the same with a shift in favor of the GPU on GB10: the ratios to sys(all) are
between 0.12 (\code{log}) and 3.4 (\code{cos}) for the real functions, exceeding one for \code{sqrt},
\code{cbrt}, \code{sin}, \code{cos}, and \code{tan}, and between 0.11 and 0.91 for the complex functions
other than \code{sqr} ($16\times$). On H100 they are between 0.15 (\code{cbrt}) and 1.5 (\code{cos}) for the
real functions, exceeding one only for \code{cos}, and between 0.16 and 0.67 for the complex functions
other than \code{sqr} ($1.6\times$).

The reason is visible in the host(1) and sys(1) columns: on the same CPU core, the converted
mini-GMP-based code is between 2.7 and 44 times slower than the system MPFR/MPC on GB10 (for example 60.8
versus 1.98\,ms for \code{sqrt} and 1835 versus 51.1\,ms for \code{log}) and between 2.3 and 37 times slower
on H100, because mini-GMP has neither the assembly kernels nor
the subquadratic multiplication and division of GMP. The parallelism of the GPU recovers a factor of
17--31 (GB10) or 14--77 (H100) over this code, which is of the same order as what was lost by replacing GMP.
For $N=4096$ function evaluations, the run-time-precision layer is therefore, on either machine, \emph{not}
a faster way of computing elementary functions than the system MPFR/MPC on all CPU cores of the same
machine. What it
provides is the evaluation of these functions, with results identical to MPFR/MPC, \emph{inside} a CUDA
kernel, where the alternative is to return to the host.

The functions differ in the amount of scratch memory they use per call, which we recorded as the peak arena
usage: at 1024 bits it ranges from 7.4\,KB (\code{sqrt}) to 198\,KB (\code{log}, \code{log1p}) for the real
functions and from 1.6\,KB (\code{sqr}) to 1.08\,MB (\code{atan}) for the complex functions. The functions
with the lowest ratios to the system libraries (\code{log}, \code{log1p}, complex \code{log}, \code{asin},
\code{acos}, \code{atan}) are those with the largest peaks; we have not separated the cost of this memory
traffic from the cost of the arithmetic. The peaks are the same on both machines.

Comparing the two GPUs, H100 evaluates every function faster than GB10 at 1024 bits, by factors between
1.1 (complex \code{sin}) and 2.9 (\code{sqrt}) with a median of 1.8. At 256 bits the factors are between
0.75 (complex \code{cos}) and 1.9 (\code{sqrt}), and GB10 is the faster machine for 10 of the 24 functions.
With 4096 active threads, the larger GPU thus has only a small advantage, in line with the dependence on
the number of active threads measured in Section~\ref{sec:analysis}.

\subsection{Basic linear algebra kernels}
\label{sec:blas}

We measured AXPY, the matrix--vector product $\mathbf{y}=A\mathbf{x}$, and the matrix--matrix product $C=AB$
with dense real and complex operands, using the run-time-precision layer at 1024 and 256 bits. These kernels
differ in how much parallelism they expose, which turns out to determine the result.

\paragraph{Implementation}
AXPY launches a fixed number of threads that stride over the elements. For the matrix--vector product we
measured two kernels. In the \emph{thread-per-row} kernel, which is the one of the previous version of this paper, one
thread computes one element of $\mathbf{y}$, that is, one sequential inner product of length $N$. In the
\emph{block-per-row} kernel, one block of 32 threads computes one element: each thread accumulates the
partial sum of every 32nd term, and the 32 partial sums are then added pairwise in a tree. In the
matrix--matrix product, one thread computes one element of $C$ by a sequential inner product. In all
kernels the accumulator is kept on the stack (\code{MPFR\_DECL\_INIT}/\code{MPC\_DECL\_INIT}) and the arena,
which then holds the temporaries of a single multiply--add, is reset at each iteration of the inner loop.
Because the tree changes the order of the additions, the system MPFR/MPC reference of the block-per-row
kernel uses the same partial sums and the same tree; all GPU results agree bit for bit with their
references. The CPU times in the tables are those of the sequential order.

\begin{table*}[t]
\centering
\caption{Linear algebra kernels of the run-time-precision layer (1024 bits), times in ms. ``Threads'' is
the number of GPU threads that perform work. host(1), sys(1), sys(all) as in Table~\ref{tab:elem_real}.}
\label{tab:linalg}
\footnotesize\setlength{\tabcolsep}{4pt}
\begin{tabular}{lrr rrrr @{\hskip 2em} rrrr}
\toprule
 & & & \multicolumn{4}{c}{GB10} & \multicolumn{4}{c}{H100} \\
\cmidrule(lr){4-7}\cmidrule(lr){8-11}
kernel & $N$ & threads & GPU & host(1) & sys(1) & sys(all) & GPU & host(1) & sys(1) & sys(all) \\
\midrule
real AXPY & 16384 & 16384 & 0.250 & 14.2 & 4.46 & 1.72 & 0.074 & 10.7 & 4.50 & 0.431 \\
complex AXPY & 16384 & 8192 & 1.03 & 58.8 & 18.7 & 2.29 & 0.428 & 42.2 & 13.6 & 1.81 \\
real mat--vec, thread/row & 256 & 256 & 5.90 & 53.4 & 17.2 & 1.96 & 7.82 & 39.5 & 11.1 & 1.68 \\
real mat--vec, block/row & 256 & 8192 & 0.899 & 53.4 & 17.2 & 1.96 & 0.365 & 39.5 & 11.1 & 1.68 \\
real mat--vec, thread/row & 1024 & 1024 & 24.1 & 852 & 275 & 16.4 & 32.2 & 594 & 169 & 19.9 \\
real mat--vec, block/row & 1024 & 32768 & 13.3 & 852 & 275 & 16.4 & 3.09 & 594 & 169 & 19.9 \\
complex mat--vec, thread/row & 128 & 128 & 16.7 & 57.0 & 18.2 & 1.97 & 19.8 & 41.5 & 11.6 & 1.42 \\
complex mat--vec, block/row & 128 & 4096 & 1.02 & 57.0 & 18.2 & 1.97 & 0.767 & 41.5 & 11.6 & 1.42 \\
complex mat--vec, thread/row & 512 & 512 & 66.8 & 948 & 291 & 16.7 & 81.2 & 643 & 180 & 21.0 \\
complex mat--vec, block/row & 512 & 16384 & 14.7 & 948 & 291 & 16.7 & 5.19 & 643 & 180 & 21.0 \\
real mat--mat & 96 & 9216 & 10.4 & 723 & 235 & 12.9 & 4.21 & 510 & 140 & 10.3 \\
real mat--mat & 192 & 16384 & 86.4 & 5786 & 1882 & 96.3 & 24.7 & 4159 & 1257 & 55.4 \\
complex mat--mat & 64 & 4096 & 17.0 & 915 & 293 & 16.4 & 12.3 & 646 & 181 & 8.12 \\
complex mat--mat & 128 & 16384 & 109 & 7319 & 2348 & 119 & 40.4 & 5163 & 1465 & 64.0 \\
\bottomrule
\end{tabular}
\end{table*}

Table~\ref{tab:linalg} shows the results at 1024 bits; the standard deviation of the GPU times was at most
4.6\% of the mean on both machines. On GB10:
\begin{itemize}
  \item \textbf{AXPY.} The GPU is $6.9\times$ (real) and $2.2\times$ (complex) faster than the system
        libraries on all cores, and $57\times$ faster than the host path on one core.
  \item \textbf{Matrix--vector product.} The thread-per-row kernel is \emph{slower} than the system
        libraries on all cores in all four cases (ratios 0.12 to 0.68). The block-per-row kernel is
        $1.8$--$6.6\times$ (real) and $4.5$--$16\times$ (complex) faster than the thread-per-row kernel and
        $1.1$--$2.2\times$ faster than the system libraries on all cores. The low figure of the previous
        version of this paper for this kernel was thus a consequence of launching only $N$ threads, each executing a
        long sequential chain, and not a property of the arithmetic or of the GPU.
  \item \textbf{Matrix--matrix product.} The GPU and the system libraries on all cores are within 25\% of
        each other (ratios 0.96 to 1.24).
\end{itemize}
At 256 bits (the raw results are distributed with the benchmark programs,
Appendix~\ref{sec:appendix_repro}) the ratios to
sys(all) are $14\times$ and $4.2\times$ for the real and complex AXPY, $1.05$--$6.6\times$ for the
block-per-row matrix--vector product, and $0.80$--$1.5\times$ for the matrix--matrix product.

As for the elementary functions, the ratios to the host path are large ($54$--$70\times$ for AXPY, the
block-per-row matrix--vector product, and the matrix--matrix product at 1024 bits), and the ratios to the
system libraries on all cores are an order of magnitude smaller. The kernels in which the run-time-precision
layer on GB10 is clearly faster than the 20 CPU cores are those with short independent work items and
little scratch memory, AXPY being the extreme case.

On H100 (ratios to the system libraries on all 32 cores):
\begin{itemize}
  \item \textbf{AXPY.} The GPU is $5.9\times$ (real) and $4.2\times$ (complex) faster at 1024 bits, and
        $4.3\times$ and $3.0\times$ faster at 256 bits.
  \item \textbf{Matrix--vector product.} The thread-per-row kernel is slower than the CPU in all four
        cases at 1024 bits (ratios 0.07 to 0.62), as on GB10. The block-per-row kernel is $10$--$26\times$
        faster than the thread-per-row kernel and $1.9$--$6.4\times$ faster than the CPU
        ($1.4$--$6.6\times$ at 256 bits).
  \item \textbf{Matrix--matrix product.} The GPU is $1.6$--$2.5\times$ faster than the CPU at 1024 bits,
        except for the smallest complex case ($N=64$, ratio 0.66); at 256 bits the ratios are between 0.55
        and 4.1.
\end{itemize}

\paragraph{Comparison of the two GPUs}
The ratios of the GPU times of the two machines depend on the kernel in the same way at both precisions.
At 1024 bits, H100 is faster than GB10 by factors of 3.4 and 2.4 for the real and complex AXPY, 1.3 to 4.3
for the block-per-row matrix--vector product, and 1.4 to 3.5 for the matrix--matrix product. For the
thread-per-row matrix--vector product, in contrast, GB10 is the faster machine in all four cases, by 16\%
to 25\% (the ratios of the GB10 time to the H100 time are 0.75 to 0.84 at 1024 bits and 0.79 to 0.85 at
256 bits). The previous version of this paper reported the same ordering and attributed it to the speed of a single
GPU thread. The thread sweep of Section~\ref{sec:analysis} supports a more precise statement: one thread
executes a multiply--add in the same time on both GPUs, H100 needs up to 26\% more time than GB10 with
$2^2$ to $2^{11}$ active threads, and H100 is faster only from $2^{12}$ threads, where the time on GB10
levels off. The thread-per-row
kernel launches at most 1024 threads and therefore falls into the range in which the smaller GPU is not at
a disadvantage; the kernels that launch $2^{13}$ threads or more benefit from the larger number of
processors of H100 (132 streaming multiprocessors against 48). We did not measure the instruction or memory
throughput of the two GPUs separately.

\subsection{Analysis of performance factors}
\label{sec:analysis}

This subsection reports the measurements made to support or refute the attributions of performance
differences to the arena, the memory layout, the per-thread state, and the number of active threads. The
register usage of the fixed-precision types was treated in Section~\ref{sec:cump}. A hardware-counter
profile could not be taken on either machine: access to the performance counters is restricted to the
administrator on both, and the profiler (\code{ncu}) reports \code{ERR\_NVGPUCTRPERM}. Statements that
would need such a profile are therefore not made.

\paragraph{Effect of the arena}
Table~\ref{tab:arena} compares three configurations of the same kernels: all allocations served by the
device \code{malloc}; the arena with a slab that holds only half of the peak usage of one work item, so
that the remaining allocations fall back to \code{malloc}; and the arena as used elsewhere in this paper.
The arena reduces the time by factors of 45 to 324 on GB10 and 241 to 542 on H100. The factor is largest
for \code{exp}, which allocates the most per call, and the half-size slab already removes between 40\% and
90\% of the time. The
device \code{malloc} is thus the dominant cost when it is used, and the arena is a precondition for every
GPU time reported in this paper.

\begin{table}[htb]
\centering
\caption{Effect of the per-thread arena on the GPU time [ms]. ``Peak'' is the peak arena usage of one work
item; ``half'' is the arena with a slab of half that size; ``ratio'' is the \code{malloc} time divided by
the arena time.}
\label{tab:arena}
\footnotesize\setlength{\tabcolsep}{1.5pt}
\begin{tabular}{llrrrrrr}
\toprule
 & kernel & bits & peak [B] & \code{malloc} & half & arena & ratio \\
\midrule
GB10 & real AXPY & 256 & 144 & 10.3 & 4.17 & 0.086 & 119 \\
GB10 & complex AXPY & 256 & 832 & 63.2 & 27.8 & 0.337 & 188 \\
GB10 & real exp & 256 & 14656 & 1001 & 251 & 3.09 & 324 \\
GB10 & real AXPY & 1024 & 432 & 11.5 & 6.81 & 0.255 & 45.2 \\
GB10 & complex AXPY & 1024 & 2464 & 94.9 & 27.0 & 1.03 & 92.6 \\
GB10 & real exp & 1024 & 60112 & 3636 & 374 & 17.6 & 207 \\
\midrule
H100 & real AXPY & 256 & 144 & 12.4 & 4.73 & 0.029 & 421 \\
H100 & complex AXPY & 256 & 832 & 89.9 & 38.3 & 0.185 & 487 \\
H100 & real exp & 256 & 14656 & 1473 & 366 & 2.72 & 542 \\
H100 & real AXPY & 1024 & 432 & 17.9 & 9.02 & 0.074 & 241 \\
H100 & complex AXPY & 1024 & 2464 & 147 & 41.4 & 0.423 & 348 \\
H100 & real exp & 1024 & 60112 & 5098 & 570 & 11.9 & 427 \\
\bottomrule
\end{tabular}
\end{table}

\paragraph{Memory layout}
In the layout of Section~\ref{sec:arena} the scratch areas of neighboring threads are one slab size apart.
Table~\ref{tab:slab} shows the time of the same AXPY kernels when only the slab size is varied. On GB10 the
time is constant up to 64\,KB and increases by 41\% (real, 16384 threads) and 24\% (complex, 8192 threads)
between 64 and 512\,KB. For the real AXPY, it then increases abruptly by a factor of 18 at 1\,MB, whereas
the complex AXPY, which launches half as many threads, shows no such step up to 2\,MB. We have not
identified the cause of this step; it does not coincide with a fixed total size of the arena (16\,GB for
the real AXPY at 1\,MB and for the complex AXPY at 2\,MB). On H100 the time does not depend on the slab
size over the whole range that could be measured (0.068--0.075\,ms for the real and 0.41--0.43\,ms for the
complex AXPY from 4\,KB to 1\,MB), so that the step is specific to GB10. The real AXPY with slabs of 2 and 4\,MB could not be run on H100: with the device stack size used
for these benchmarks, about 25\,GB of device memory remained free, less than the 32 and 64\,GB of these
arenas. The slab should in any case be chosen close to
the peak usage of the kernel, which the library reports. We did not measure a layout in which the limbs of
different threads are interleaved, and without a profile we make no statement on whether memory access is
the main bottleneck of the run-time-precision layer.

\begin{table}[htb]
\centering
\caption{GPU time [ms] of the real (16384 threads) and complex (8192 threads) AXPY of the
run-time-precision layer as a function of the slab size ($N=16384$, 1024 bits). --: not measured (complex,
4\,MB) or not allocatable (H100, real; see the text).}
\label{tab:slab}
\footnotesize\setlength{\tabcolsep}{4pt}
\begin{tabular}{rrrrr}
\toprule
 & \multicolumn{2}{c}{GB10} & \multicolumn{2}{c}{H100} \\
\cmidrule(lr){2-3}\cmidrule(lr){4-5}
slab [KB] & real & complex & real & complex \\
\midrule
4 & 0.246 & 0.870 & 0.068 & 0.414 \\
8 & 0.251 & 0.878 & 0.068 & 0.416 \\
16 & 0.246 & 0.886 & 0.068 & 0.416 \\
32 & 0.246 & 0.877 & 0.069 & 0.415 \\
64 & 0.252 & 0.878 & 0.070 & 0.422 \\
128 & 0.287 & 0.933 & 0.070 & 0.416 \\
256 & 0.313 & 1.03 & 0.070 & 0.431 \\
512 & 0.355 & 1.09 & 0.070 & 0.425 \\
1024 & 6.37 & 1.10 & 0.075 & 0.425 \\
2048 & 8.00 & 1.13 & -- & 0.417 \\
4096 & 8.42 & -- & -- & -- \\
\bottomrule
\end{tabular}
\end{table}

\paragraph{Cost of the per-thread state}
We timed the real and complex AXPY and the real \code{exp} with the per-thread state blocks of
Section~\ref{sec:arena} and with the single shared state of the previous version. On GB10 the difference
is between $-0.2\%$ and $+3.7\%$ at 1024 bits and between $0.0\%$ and $+6.4\%$ at 256 bits, the largest
value being that of the real AXPY at 256 bits (0.086 versus 0.081\,ms), whose standard deviation is 3\% of
the mean. On H100 the difference is between $-1.9\%$ and $+1.9\%$. The separation of the state costs at
most a few percent.

\paragraph{Number of active threads}
We use ``number of active threads'' for the number of GPU threads that perform work in a kernel. This is
different from \emph{occupancy} in the CUDA sense, which we have not measured. Table~\ref{tab:threads} shows
the time of the real AXPY with $N=65536$ when 1 to $2^{16}$ threads are launched. On GB10 the time is
inversely proportional to the number of threads up to $2^{11}$ threads (a factor of 1310 at 256 bits and
1010 at 1024 bits relative to one thread), is shortest at $2^{13}$ threads, and does not decrease further.
A single GPU thread needs 8.8\,$\mu$s (256 bits) and 17.4\,$\mu$s (1024 bits) per multiply--add, which is
122 and 64 times the time per element of the system MPFR on one CPU core in Table~\ref{tab:linalg} and its
256-bit counterpart. On H100 a single thread needs the same time (8.8 and 17.7\,$\mu$s; 84 and 64 times the
time of one CPU core), and H100 needs up to 26\% more time than GB10 between $2^2$ and $2^{11}$ threads, but
H100 continues to scale up to $2^{15}$ threads, where it is 3.8 (256 bits) and 4.0 (1024 bits) times faster than
GB10. On either machine, the run-time-precision layer therefore needs a few thousand active threads before it
can match a multithreaded CPU, which is the situation of AXPY, the block-per-row matrix--vector product,
and the matrix--matrix product, but not of the thread-per-row matrix--vector product.

\begin{table}[htb]
\centering
\caption{GPU time [ms] of the real AXPY of the run-time-precision layer ($N=65536$) as a function of the
number of launched threads.}
\label{tab:threads}
\footnotesize\setlength{\tabcolsep}{4pt}
\begin{tabular}{rrrrr}
\toprule
 & \multicolumn{2}{c}{GB10} & \multicolumn{2}{c}{H100} \\
\cmidrule(lr){2-3}\cmidrule(lr){4-5}
threads & 256 bits & 1024 bits & 256 bits & 1024 bits \\
\midrule
$2^{0}$ & 575 & 1140 & 576 & 1161 \\
$2^{1}$ & 323 & 683 & 324 & 705 \\
$2^{2}$ & 178 & 398 & 194 & 425 \\
$2^{3}$ & 96.2 & 219 & 99.7 & 231 \\
$2^{4}$ & 51.2 & 117 & 56.7 & 127 \\
$2^{5}$ & 27.8 & 67.0 & 33.3 & 80.8 \\
$2^{6}$ & 14.6 & 37.1 & 17.0 & 40.3 \\
$2^{7}$ & 7.36 & 18.1 & 8.46 & 19.9 \\
$2^{8}$ & 3.68 & 9.08 & 4.23 & 10.1 \\
$2^{9}$ & 1.87 & 4.57 & 2.08 & 5.09 \\
$2^{10}$ & 0.893 & 2.23 & 1.09 & 2.57 \\
$2^{11}$ & 0.438 & 1.13 & 0.551 & 1.30 \\
$2^{12}$ & 0.320 & 0.982 & 0.280 & 0.651 \\
$2^{13}$ & 0.262 & 0.785 & 0.143 & 0.350 \\
$2^{14}$ & 0.317 & 0.976 & 0.092 & 0.241 \\
$2^{15}$ & 0.288 & 0.877 & 0.075 & 0.217 \\
$2^{16}$ & 0.300 & 0.890 & 0.076 & 0.231 \\
\bottomrule
\end{tabular}
\end{table}

\section{Application: simultaneous polynomial root-finding}
\label{sec:rootfinding}

As an application in which high precision is required by the problem itself, we compute all roots of a
polynomial $P(z)=a_0 z^n + a_1 z^{n-1} + \dots + a_n$ ($a_0\neq0$) by simultaneous iteration. All $n$ root
approximations are updated in each iteration, which provides $n$-fold parallelism, and the two test problems
below are out of reach of these methods in binary64 arithmetic at the degrees considered.

\subsection{Problems and input data}
\label{sec:root_problems}

\begin{description}
  \item[Wilkinson polynomial]
\begin{equation}
  P_n(z) = \prod_{k=1}^{n} (z-k) .
\end{equation}
The roots are the integers $1,2,\dots,n$, the coefficients grow to approximately $n!$ in magnitude, and the
roots are extremely sensitive to perturbations of the coefficients. The coefficients are integers and are
generated exactly by repeated multiplication by $(z-k)$. Their magnitude is of the order of $n!$
($\log_2 n!\approx296$ for $n=64$ and 716 for $n=128$); since they have trailing zero bits, the longest
significand among the coefficients, computed from the exact integers, is somewhat shorter: 107, 265, 446,
641, 846, and 1059 bits for $n=32$, 64, 96, 128, 160, and 192.

\item[Chebyshev quadrature polynomial]
The polynomial that arises from the moment conditions of equal-weight (Chebyshev) quadrature on $[-1,1]$,
whose roots are the nodes of that rule; it is not the Chebyshev polynomial $T_n$, and at the degrees
considered here its roots include complex values. We refer to it as the Chebyshev problem and to its roots
as Chebyshev nodes. With
$a_0=1$, the coefficients are generated by the recurrence
\begin{equation}
  a_{2i} \;=\; -\frac{n}{2i}\sum_{j=1}^{i}\frac{1}{2j+1}\,a_{2(i-j)},\ (i=1,\dots,\lfloor n/2\rfloor),
\end{equation}
with $a_{2i-1}=0$. The coefficients are rational numbers of order one, and the roots are a mixture of real
and complex roots inside the unit disk.
\end{description}

\paragraph{The polynomial actually solved}
The coefficients are generated once, exactly, on the host: integers (GMP \code{mpz}) for the Wilkinson
polynomial and rational numbers (\code{mpq}) for the recurrence of the Chebyshev nodes. They are then
rounded once to the working precision \PB{} (round to nearest), and the resulting bit patterns are passed to
all three implementations of Section~\ref{sec:root_impl}; each implementation reads the coefficients back
from its own storage, and the program checks that the three copies are identical. In the previous version
the Chebyshev coefficients had been generated separately on the CPU and on the GPU and could differ in the
last bits.

Two kinds of problem must therefore be distinguished.
\begin{itemize}
  \item \emph{Exact-coefficient problem.} If no coefficient is changed by the rounding, the polynomial
        that is solved is the mathematical one. This holds for the Wilkinson polynomial when \PB{} is at
        least the length of the longest significand among its coefficients (the values given above). In the
        convergence experiments this is the case for $n=32$ at all three precisions, for $n=64$ and
        $n=96$ from 512 bits, and for $n=128$ and $n=160$ at 1024 bits.
  \item \emph{Rounded-coefficient problem.} Otherwise the polynomial that is solved is a nearby polynomial
        $\tilde P$. This is the case for the Wilkinson polynomial when \PB{} is smaller than that length, whose roots can
        then differ substantially from $1,\dots,n$, and for all Chebyshev cases, whose coefficients are
        not dyadic.
\end{itemize}
All errors reported below are measured against the roots of the polynomial that was actually solved:
the integers in the first case, and in the second case reference roots of $\tilde P$ computed with the
system MPC at $\max(4\PB,2048)$ bits or more and accepted only if
$n\,|\tilde P(\xi)/\tilde P'(\xi)|$, which in exact arithmetic bounds the distance from $\xi$ to the nearest
root of $\tilde P$, is below $2^{-\PB-32}$ relative to $\max(1,|\xi|)$. This quantity is evaluated in
floating-point arithmetic at the reference precision and is therefore an estimate, not a rigorous
enclosure; that the reference roots are distinct and are matched one-to-one with the computed roots is
checked separately.

\subsection{Methods and implementations}
\label{sec:root_impl}

Let $x_1,\dots,x_n$ be the current approximations. The initial values are placed on a circle whose center and
radius are defined for each problem, at the angles $2\pi i/n+\pi/(2n)$, $i=0,\dots,n-1$; they are computed
once with MPFR at \PB{} bits and passed to the three implementations in the same way as the coefficients.

\begin{description}
\item[Higher-order Ehrlich--Aberth method]
With $T_i^{(0)}=x_i$,
\begin{equation}
  T_i^{(\ell+1)} = x_i - \frac{P(x_i)}{\,P'(x_i) - P(x_i)\sum_{j\neq i}\frac{1}{x_i - T_j^{(\ell)}}\,}
\end{equation}
for $\ell=0,\dots,L-1$, followed by $x_i \leftarrow T_i^{(L)}$. $L=1$ is the classical Ehrlich--Aberth method
of order 3, and the formal order is $2L+1$ in general. We use $L=3$ (formal order 7).

\item[Sakurai--Torii--Sugiura (STS) method~\cite{sakurai_sts1991}]
Around each approximation $x_k$, $f(w)=P(x_k+w)$ and $g_k(w)=\prod_{i\neq k}\bigl((x_k-z^\ast_i)+w\bigr)$ are
expanded in Taylor series up to degree $m$, and $x_k$ is updated from the coefficients of the quotient
series $q=g_k/f$ by
\begin{equation}
  x_k \leftarrow x_k + \frac{q_{m-1}}{q_m}
\end{equation}
(Algorithm A). Algorithm A$'$ first refines the other approximations $z^\ast$ and then applies A (formal
order $2m+1$). We use $m=4$ and Algorithm A$'$ (formal order 9).
\end{description}

Three implementations are built into one program and run on the same rounded coefficients and the same
initial values.
\begin{itemize}
  \item \textbf{CPU}: GMP 6.3.0, MPFR 4.2.2, and MPC 1.4.1 at run-time precision; the loop over the
        roots is parallelized with OpenMP.
  \item \textbf{GPU \texttt{cu\_mpc}}: the run-time-precision layer. One root is assigned to one thread
        block; the evaluation of $P$ and $P'$ by Horner's rule is divided into segments within the block,
        and the segments and the repulsion sum $\sum_{j\neq i}1/(x_i-T_j)$ are combined by a tree
        reduction. Each operation uses the per-thread arena and the per-thread state.
  \item \textbf{GPU \texttt{cu\_fcomplex}}: the fixed-precision layer, without arena. The Ehrlich--Aberth
        method uses the same one-block-per-root structure; the STS method uses one \emph{thread} per root.
        Of the operations of this layer, the programs use $+,-,\times$, and the complex division.
\end{itemize}
The precision is restricted to 128, 256, 512, and 1024 bits, the values instantiated for \fcomplex. The
stopping rule, the accuracy measures, and the timing are the same host code for the three implementations.

\paragraph{CPU configuration}
The time of the CPU implementation depends on the number and placement of the OpenMP threads, and the best
choice depends on the problem size. On GB10, whose CPU has 10 Cortex-X925 and 10 Cortex-A725 cores, we
measured three configurations: 20 threads without binding, 20 threads bound to cores, and 10 threads bound
to the Cortex-X925 cores. On H100, whose host has two 16-core processors, we measured five: 20 threads
without binding (the configuration of the previous version), 16 and 32 threads with \code{close} binding,
and 20 and 32 threads with \code{spread} binding. For every case, the CPU time used below is that of the \emph{fastest} configuration for that
case.

\paragraph{What can and cannot be compared}
The three implementations do not produce bit-identical iterates, for two reasons that are independent of the
correctness of the arithmetic: the tree reductions on the GPU add and multiply in a different order from the
sequential loops on the CPU, and the division of \fcomplex{} is not correctly rounded
(Section~\ref{sec:verification}). The iterates agree only up to rounding errors of the working precision.
We evaluate throughput and time to solution separately:
\begin{itemize}
  \item \textbf{Throughput}: the number of iterations is fixed (Ehrlich--Aberth: 20, STS: 12), and
        \begin{equation}
          \text{speedup} \;=\; \frac{\text{CPU time}}{\text{GPU time}}
        \end{equation}
        is the ratio of the times per iteration, each obtained from the minimum elapsed time over five
        runs. It says nothing about the
        time to obtain a solution.
  \item \textbf{Convergence and time to solution}: each solver is run until the largest correction falls
        below $\text{tol}=2^{-\lfloor 0.75\,\PB\rfloor}$ ($2^{-192}$ at 256 bits, $2^{-768}$ at 1024 bits),
        up to 800 (Ehrlich--Aberth) or 400 (STS) iterations. A run is also stopped, and counted as
        \emph{not converged}, when the smallest value reached so far by the largest correction of an
        iteration has not halved during the last 100 iterations (stagnation). The time is the wall-clock time of the whole iteration loop, which on
        the GPU includes the transfer of the $n$ corrections to the host in every iteration.
\end{itemize}
For every run we report two accuracy measures, both evaluated on the host with MPFR at $2\PB+64$ bits from
the full-precision roots: the \emph{forward error} $\max_i\min_j|x_i-\xi_j|$ against the reference roots
$\xi_j$ defined in Section~\ref{sec:root_problems} (we also check that the assignment of computed roots to
reference roots is one-to-one), and the \emph{normalized residual}
\begin{equation}
  \max_i \frac{|\tilde P(x_i)|}{\sum_{k=0}^{n}|\tilde a_k|\,|x_i|^{n-k}} ,
\end{equation}
which is of the order of the unit roundoff $2^{-\PB}$ when $x_i$ is a root of a polynomial whose
coefficients are within rounding error of the $\tilde a_k$.

\subsection{Convergence and accuracy}
\label{sec:root_conv}

Tables~\ref{tab:root_conv} and~\ref{tab:root_conv_sts} show, for every degree and precision, the number of
iterations of the three implementations and the two accuracy measures of the \texttt{cu\_fcomplex}
implementation. The tables are those of GB10; on H100 the iteration counts, the stopping reasons, the
forward errors, and the normalized residuals of all 216 runs (two methods, 36 cases, three implementations)
are identical to those on GB10 in every printed digit, so that everything in this subsection holds for both
machines.

\begin{table*}[t]
\centering
\caption{Higher-order Ehrlich--Aberth method (identical on GB10 and H100): iterations, forward error, and normalized residual.
A single number in the column ``it'' means that the CPU, \texttt{cu\_mpc}, and \texttt{cu\_fcomplex}
implementations needed the same number of iterations; otherwise the three numbers are given in this order.
$^{+}$: stopped by the stagnation rule; $^{*}$: iteration limit reached (not converged in both cases).
The errors are those of \texttt{cu\_fcomplex}.}
\label{tab:root_conv}
\footnotesize\setlength{\tabcolsep}{1.6pt}
\begin{tabular}{lr lll lll lll}
\toprule
 & & \multicolumn{3}{c}{256 bits} & \multicolumn{3}{c}{512 bits} & \multicolumn{3}{c}{1024 bits} \\
\cmidrule(lr){3-5}\cmidrule(lr){6-8}\cmidrule(lr){9-11}
problem & $n$ & it & fwd.\ err. & norm.\ res. & it & fwd.\ err. & norm.\ res. & it & fwd.\ err. & norm.\ res. \\
\midrule
\multirow{6}{*}{Wilkinson} & 32 & 166$^{+}$/113$^{+}$/113$^{+}$ & 1.9\mbox{e}-54 & 3.6\mbox{e}-78 & 14 & 4.3\mbox{e}-131 & 3.6\mbox{e}-155 & 14 & 1.7\mbox{e}-285 & 2.0\mbox{e}-309 \\
 & 64 & 144$^{+}$/134$^{+}$/134$^{+}$ & 8.1\mbox{e}-30 & 4.5\mbox{e}-78 & 123$^{+}$/126$^{+}$/126$^{+}$ & 9.9\mbox{e}-107 & 3.6\mbox{e}-155 & 24 & 7.8\mbox{e}-261 & 4.2\mbox{e}-309 \\
 & 96 & 156$^{+}$/349$^{+}$/349$^{+}$ & 2.0\mbox{e}-5 & 3.9\mbox{e}-78 & 144$^{+}$/136$^{+}$/136$^{+}$ & 1.7\mbox{e}-82 & 2.5\mbox{e}-155 & 36 & 2.8\mbox{e}-236 & 2.5\mbox{e}-309 \\
 & 128 & 208$^{+}$/257$^{+}$/257$^{+}$ & 9.5\mbox{e}0 & 3.1\mbox{e}-78 & 149$^{+}$/155$^{+}$/155$^{+}$ & 4.0\mbox{e}-58 & 2.5\mbox{e}-155 & 148$^{+}$ & 2.9\mbox{e}-212 & 2.0\mbox{e}-309 \\
 & 160 & 129$^{+}$/117$^{+}$/117$^{+}$ & 2.1\mbox{e}1 & 9.5\mbox{e}-71 & 196$^{+}$/413$^{+}$/413$^{+}$ & 1.7\mbox{e}-33 & 3.1\mbox{e}-155 & 155$^{+}$ & 9.5\mbox{e}-188 & 2.0\mbox{e}-309 \\
 & 192 & 101$^{+}$/102$^{+}$/102$^{+}$ & 4.6\mbox{e}1 & 2.3\mbox{e}-74 & 177$^{+}$/188$^{+}$/188$^{+}$ & 5.9\mbox{e}-9 & 2.2\mbox{e}-155 & 236$^{+}$/166$^{+}$/166$^{+}$ & 3.1\mbox{e}-163 & 1.9\mbox{e}-309 \\
\midrule
\multirow{6}{*}{Chebyshev} & 32 & 9 & 5.5\mbox{e}-73 & 1.9\mbox{e}-77 & 9 & 6.9\mbox{e}-150 & 1.9\mbox{e}-154 & 10 & 4.1\mbox{e}-304 & 2.3\mbox{e}-308 \\
 & 64 & 15 & 1.7\mbox{e}-67 & 7.8\mbox{e}-77 & 15 & 2.7\mbox{e}-144 & 1.8\mbox{e}-154 & 15 & 1.6\mbox{e}-298 & 2.3\mbox{e}-308 \\
 & 96 & 19 & 3.2\mbox{e}-61 & 7.1\mbox{e}-77 & 19 & 2.2\mbox{e}-138 & 9.3\mbox{e}-154 & 20 & 3.1\mbox{e}-292 & 2.3\mbox{e}-308 \\
 & 128 & 125$^{+}$ & 1.5\mbox{e}-55 & 1.2\mbox{e}-76 & 25 & 2.3\mbox{e}-132 & 8.8\mbox{e}-154 & 25 & 1.4\mbox{e}-286 & 6.0\mbox{e}-308 \\
 & 160 & 151$^{+}$/129$^{+}$/129$^{+}$ & 2.6\mbox{e}-49 & 3.7\mbox{e}-77 & 30 & 2.6\mbox{e}-126 & 1.5\mbox{e}-153 & 30 & 2.8\mbox{e}-280 & 9.8\mbox{e}-308 \\
 & 192 & 229$^{+}$/135$^{+}$/135$^{+}$ & 2.4\mbox{e}-43 & 1.3\mbox{e}-76 & 35 & 4.0\mbox{e}-120 & 2.1\mbox{e}-153 & 35 & 1.4\mbox{e}-274 & 1.4\mbox{e}-307 \\
\bottomrule
\end{tabular}
\end{table*}

\begin{table*}[t]
\centering
\caption{STS method (identical on GB10 and H100): iterations, forward error, and normalized residual. Notation as in
Table~\ref{tab:root_conv}.}
\label{tab:root_conv_sts}
\footnotesize\setlength{\tabcolsep}{1.6pt}
\begin{tabular}{lr lll lll lll}
\toprule
 & & \multicolumn{3}{c}{256 bits} & \multicolumn{3}{c}{512 bits} & \multicolumn{3}{c}{1024 bits} \\
\cmidrule(lr){3-5}\cmidrule(lr){6-8}\cmidrule(lr){9-11}
problem & $n$ & it & fwd.\ err. & norm.\ res. & it & fwd.\ err. & norm.\ res. & it & fwd.\ err. & norm.\ res. \\
\midrule
\multirow{6}{*}{Wilkinson} & 32 & 134$^{+}$/117$^{+}$/134$^{+}$ & 3.6\mbox{e}-55 & 2.4\mbox{e}-78 & 16 & 8.4\mbox{e}-132 & 1.3\mbox{e}-155 & 17 & 5.5\mbox{e}-286 & 1.2\mbox{e}-309 \\
 & 64 & 205$^{+}$/130$^{+}$/205$^{+}$ & 3.2\mbox{e}-30 & 1.9\mbox{e}-78 & 127$^{+}$ & 2.2\mbox{e}-107 & 1.3\mbox{e}-155 & 27 & 1.7\mbox{e}-261 & 1.0\mbox{e}-309 \\
 & 96 & 199$^{+}$/400$^{*}$/199$^{+}$ & 1.2\mbox{e}-5 & 1.1\mbox{e}-78 & 140$^{+}$ & 5.9\mbox{e}-83 & 1.1\mbox{e}-155 & 41 & 8.4\mbox{e}-237 & 1.0\mbox{e}-309 \\
 & 128 & 101$^{+}$/400$^{*}$/101$^{+}$ & 3.0\mbox{e}0 & 1.8\mbox{e}-78 & 248$^{+}$/157$^{+}$/248$^{+}$ & 2.9\mbox{e}-58 & 1.5\mbox{e}-155 & 151$^{+}$ & 2.5\mbox{e}-212 & 9.3\mbox{e}-310 \\
 & 160 & 143$^{+}$/106$^{+}$/143$^{+}$ & 3.7\mbox{e}0 & 1.0\mbox{e}-78 & 167$^{+}$/327$^{+}$/167$^{+}$ & 5.4\mbox{e}-34 & 1.5\mbox{e}-155 & 265$^{+}$/169$^{+}$/265$^{+}$ & 4.3\mbox{e}-188 & 7.6\mbox{e}-310 \\
 & 192 & 144$^{+}$/103$^{+}$/144$^{+}$ & 1.6\mbox{e}0 & 1.3\mbox{e}-78 & 206$^{+}$/204$^{+}$/206$^{+}$ & 2.2\mbox{e}-9 & 1.0\mbox{e}-155 & 181$^{+}$ & 1.4\mbox{e}-163 & 9.1\mbox{e}-310 \\
\midrule
\multirow{6}{*}{Chebyshev} & 32 & 10 & 4.4\mbox{e}-73 & 1.8\mbox{e}-77 & 11 & 2.1\mbox{e}-150 & 2.0\mbox{e}-154 & 11 & 2.1\mbox{e}-304 & 2.3\mbox{e}-308 \\
 & 64 & 16 & 1.9\mbox{e}-67 & 3.3\mbox{e}-77 & 17 & 1.8\mbox{e}-144 & 1.8\mbox{e}-154 & 17 & 7.1\mbox{e}-299 & 2.0\mbox{e}-308 \\
 & 96 & 22 & 9.7\mbox{e}-62 & 7.1\mbox{e}-77 & 22 & 1.0\mbox{e}-138 & 6.5\mbox{e}-154 & 22 & 8.8\mbox{e}-293 & 2.2\mbox{e}-308 \\
 & 128 & 164$^{+}$/129$^{+}$/164$^{+}$ & 1.4\mbox{e}-55 & 1.1\mbox{e}-76 & 27 & 1.5\mbox{e}-132 & 8.2\mbox{e}-154 & 27 & 1.1\mbox{e}-286 & 5.9\mbox{e}-308 \\
 & 160 & 168$^{+}$/134$^{+}$/168$^{+}$ & 1.4\mbox{e}-49 & 3.7\mbox{e}-77 & 32 & 7.1\mbox{e}-127 & 9.6\mbox{e}-154 & 32 & 1.1\mbox{e}-280 & 9.4\mbox{e}-308 \\
 & 192 & 175$^{+}$/139$^{+}$/175$^{+}$ & 2.1\mbox{e}-43 & 1.2\mbox{e}-76 & 35 & 1.2\mbox{e}-120 & 1.0\mbox{e}-153 & 35 & 5.7\mbox{e}-275 & 9.3\mbox{e}-308 \\
\bottomrule
\end{tabular}
\end{table*}

\paragraph{Agreement of the three implementations}
Of the 36 cases of each method (two problems, $n\in\{32,64,96,128,160,192\}$, $\PB\in\{256,512,1024\}$), all three implementations converge in 19 and none converges in the other
17; there is no case in which only some of them converge. In all 38 converged cases the three
implementations need \emph{exactly the same number of iterations}, and their forward errors agree within a
factor of 5.1. The differences in rounding between the implementations (order of summation, division of
\fcomplex) thus did not affect the observed iteration counts, although the forward errors differ by up to
this factor. The previous version of this paper reported that the
\texttt{cu\_fcomplex} implementation often needed fewer iterations than the CPU implementation. That
difference was an artifact of the initial values, which were placed symmetrically with respect to the real
axis (angles $2\pi(i+\frac12)/n$). With such conjugate-symmetric initial values, convergence was much
slower and sensitive to rounding-level differences. This behavior is consistent with a restriction imposed
by the symmetry: for a polynomial with real coefficients, a conjugate pair of approximations cannot
converge to two distinct real roots while the symmetry is preserved exactly, and both test polynomials
have real roots. We did not analyze the mechanism further.
Table~\ref{tab:root_sym} shows the observations for the cases that converge. With the symmetric initial values the
Ehrlich--Aberth method needs 87--371 iterations, 4.7 to 37 times as many as with the asymmetric initial
values used in this paper, and the number increases with the precision. The counts of the CPU implementation and of the
two GPU implementations then differ in all 19 cases, by up to 3 iterations; for the STS method the counts
differ in one of the 19 cases. The dependence on rounding details is extreme: when the CPU implementation
was linked with MPC~1.4.0 instead of 1.4.1, with everything else unchanged, it needed 27--48 iterations in
the same 19 Ehrlich--Aberth cases, whereas the results with the asymmetric initial values did not change.
With the asymmetric initial values the counts coincide and are smaller. Iteration counts obtained with
symmetric initial values are therefore not used to compare implementations.

\begin{table}[htb]
\centering
\caption{Iterations to convergence with initial values that are symmetric with respect to the real axis
(as in the previous version of this paper), compared with the asymmetric initial values used in this paper
($n\in\{32,96,192\}$; identical on GB10 and H100).}
\label{tab:root_sym}
\footnotesize\setlength{\tabcolsep}{3pt}
\begin{tabular}{llrr rrr r}
\toprule
 & & & & \multicolumn{3}{c}{symmetric} & asym. \\
\cmidrule(lr){5-7}
 & problem & $n$ & bits & CPU & \texttt{cu\_mpc} & \texttt{cu\_fc} & all \\
\midrule
E--A & Chebyshev & 32 & 256 & 96 & 98 & 98 & 9 \\
E--A & Chebyshev & 32 & 512 & 188 & 187 & 187 & 9 \\
E--A & Chebyshev & 32 & 1024 & 371 & 370 & 370 & 10 \\
E--A & Chebyshev & 96 & 256 & 87 & 90 & 90 & 19 \\
E--A & Chebyshev & 96 & 512 & 181 & 180 & 180 & 19 \\
E--A & Chebyshev & 96 & 1024 & 363 & 364 & 364 & 20 \\
E--A & Chebyshev & 192 & 512 & 169 & 168 & 168 & 35 \\
E--A & Chebyshev & 192 & 1024 & 351 & 352 & 352 & 35 \\
E--A & Wilkinson & 32 & 512 & 172 & 173 & 173 & 14 \\
E--A & Wilkinson & 32 & 1024 & 353 & 356 & 356 & 14 \\
E--A & Wilkinson & 96 & 1024 & 314 & 313 & 313 & 36 \\
STS & Chebyshev & 32 & 256 & 15 & 15 & 15 & 10 \\
STS & Chebyshev & 32 & 512 & 23 & 23 & 23 & 11 \\
STS & Chebyshev & 32 & 1024 & 38 & 38 & 38 & 11 \\
STS & Chebyshev & 96 & 256 & 22 & 22 & 22 & 22 \\
STS & Chebyshev & 96 & 512 & 27 & 27 & 27 & 22 \\
STS & Chebyshev & 96 & 1024 & 39 & 39 & 39 & 22 \\
STS & Chebyshev & 192 & 512 & 36 & 36 & 36 & 35 \\
STS & Chebyshev & 192 & 1024 & 47 & 47 & 47 & 35 \\
STS & Wilkinson & 32 & 512 & 145 & 145 & 145 & 16 \\
STS & Wilkinson & 32 & 1024 & 298 & 298 & 298 & 17 \\
STS & Wilkinson & 96 & 1024 & 263 & 263 & 263 & 41 \\
\bottomrule
\end{tabular}
\end{table}

\paragraph{Chebyshev quadrature polynomial}
All runs converge at 512 and 1024 bits, and at 256 bits for $n\le96$. The forward error is far larger than
the unit roundoff and grows with the degree---at 256 bits from $5.5\times10^{-73}$ for $n=32$ to
$2.4\times10^{-43}$ for $n=192$---while the normalized residual stays within a factor of 30 of $2^{-\PB}$
in every case. Each computed root, taken by itself, is thus an exact root of some polynomial within
rounding error of $\tilde P$ (a small backward error for each root separately; the measure does not show
that all computed roots are roots of one common nearby polynomial). This is consistent with forward errors
that are determined by the conditioning of the roots, which we did not compute, and not by the iteration. The three
runs at 256 bits that do not meet the criterion ($n\ge128$) are of this kind: their corrections stagnate
at $10^{-55}$ to $10^{-43}$, which is the level of their forward errors ($1.5\times10^{-55}$ to
$2.4\times10^{-43}$) and above the tolerance $2^{-192}\approx1.6\times10^{-58}$, while their normalized
residuals ($10^{-76}$) are as small as those of the converged runs. The observed stagnation is consistent with the amplification of rounding errors by ill-conditioned
roots, in which case the tolerance $2^{-0.75\PB}$ is below what can be reached at \PB{} bits; a criterion
based on the normalized residual would have accepted these runs.

\paragraph{Wilkinson polynomial}
The same holds, more strongly, for the Wilkinson polynomial. The criterion is met only for
$(n,\PB)=(32,512)$, $(32,1024)$, $(64,1024)$, and $(96,1024)$. In the other cases with exact coefficients
(for example $n=128$ at 1024 bits: forward error $2.9\times10^{-212}$, tolerance $6.4\times10^{-232}$,
normalized residual $2.0\times10^{-309}$), and in the rounded-coefficient cases at 512 and 1024 bits, the
corrections stagnate while the normalized residual is at the level of $2^{-\PB}$: the computed roots have
small backward errors but are less accurate than the tolerance demands, as is to be expected for roots as
ill-conditioned as these.
At 256 bits and $n\ge128$ the situation is different: the forward errors are of order 1 to 10, and the
computed roots cannot even be assigned one-to-one to the roots of $\tilde P$. At this precision the roots
are not resolved at all, and these runs are meaningful only as throughput measurements. In no case is the
error with respect to the integers $1,\dots,n$ reported for a rounded-coefficient problem.

\subsection{Throughput per iteration}
\label{sec:root_throughput}

\begin{figure*}[t]\centering
  \includegraphics[width=\linewidth]{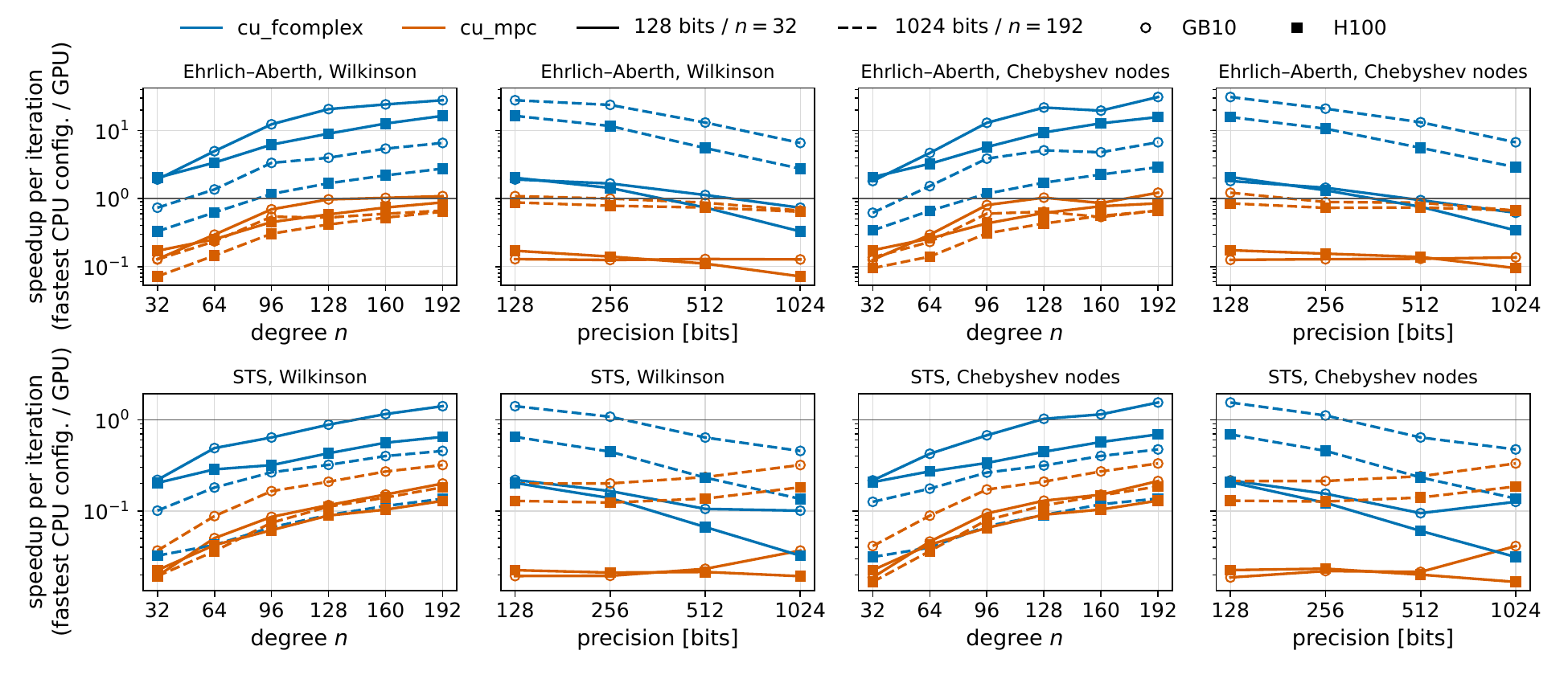}
  \caption{Speedup per iteration of the two GPU implementations over the fastest CPU configuration
  of the same machine (${>}1$: the GPU is faster). For each method and problem, the left panel varies the
  degree at 128 and 1024 bits and the right panel varies the precision at $n=32$ and $n=192$. Open
  circles: GB10; filled squares: H100.}
  \label{fig:root_speedup}
\end{figure*}

Figure~\ref{fig:root_speedup} shows the speedup per iteration over the fastest CPU configuration of each
machine, and Table~\ref{tab:root_corner} lists the values at the four corners of the parameter range.

\begin{table*}[t]
\centering
\caption{Speedup per iteration over the fastest CPU configuration at the four corners $(n,\PB)$ of the
parameter range. E--A: higher-order Ehrlich--Aberth.}
\label{tab:root_corner}
\footnotesize\setlength{\tabcolsep}{5pt}
\begin{tabular}{lll rrrr rrrr}
\toprule
 & & & \multicolumn{4}{c}{GB10} & \multicolumn{4}{c}{H100} \\
\cmidrule(lr){4-7}\cmidrule(lr){8-11}
 & & & \multicolumn{2}{c}{128 bits} & \multicolumn{2}{c}{1024 bits} & \multicolumn{2}{c}{128 bits} & \multicolumn{2}{c}{1024 bits} \\
 & problem & GPU & 32 & 192 & 32 & 192 & 32 & 192 & 32 & 192 \\
\midrule
E--A & Wilkinson & \texttt{cu\_mpc} & 0.13 & 1.09 & 0.13 & 0.67 & 0.17 & 0.88 & 0.07 & 0.64 \\
E--A & Wilkinson & \texttt{cu\_fcomplex} & 1.91 & 27.90 & 0.73 & 6.58 & 2.02 & 16.45 & 0.33 & 2.76 \\
E--A & Chebyshev & \texttt{cu\_mpc} & 0.13 & 1.22 & 0.14 & 0.67 & 0.17 & 0.85 & 0.10 & 0.67 \\
E--A & Chebyshev & \texttt{cu\_fcomplex} & 1.81 & 31.12 & 0.62 & 6.73 & 2.08 & 15.79 & 0.34 & 2.90 \\
\midrule
STS & Wilkinson & \texttt{cu\_mpc} & 0.02 & 0.20 & 0.04 & 0.32 & 0.02 & 0.13 & 0.02 & 0.18 \\
STS & Wilkinson & \texttt{cu\_fcomplex} & 0.22 & 1.41 & 0.10 & 0.46 & 0.20 & 0.65 & 0.03 & 0.14 \\
STS & Chebyshev & \texttt{cu\_mpc} & 0.02 & 0.21 & 0.04 & 0.33 & 0.02 & 0.13 & 0.02 & 0.19 \\
STS & Chebyshev & \texttt{cu\_fcomplex} & 0.22 & 1.55 & 0.13 & 0.47 & 0.21 & 0.69 & 0.03 & 0.14 \\
\bottomrule
\end{tabular}
\end{table*}

The throughput grid consists of 42 cases per method (two problems; $n\in\{32,64,96,128,160,192\}$ at 128,
256, and 1024 bits, and $n\in\{32,96,192\}$ at 512 bits).
\begin{itemize}
  \item \textbf{Ehrlich--Aberth, \texttt{cu\_fcomplex}.} The GPU is faster than the fastest CPU
        configuration in 39 of the 42 cases on GB10 and in 36 on H100. The speedup is largest at the
        highest degree and lowest precision, $(n,\PB)=(192,128)$: $28$--$31\times$ on GB10 and
        $16\times$ on H100. It decreases with the precision and with decreasing degree: at $(192,1024)$
        it is $6.6$--$6.7\times$ on GB10 and $2.8$--$2.9\times$ on H100, at $(32,128)$ $1.8$--$2.1\times$
        on both machines, and at $(32,1024)$ the GPU is slower ($0.62$--$0.73\times$ and
        $0.33$--$0.34\times$). The cases below one are $n=32$ at 1024 bits and, for the Chebyshev
        problem, at 512 bits on GB10, and $n=32$ at 512 and 1024 bits and $n=64$ at 1024 bits on H100.
  \item \textbf{Ehrlich--Aberth, \texttt{cu\_mpc}.} The run-time-precision layer is slower than the CPU in
        37 of the 42 cases on GB10 and in all 42 on H100; the speedup ranges from $0.13\times$ to
        $1.22\times$ on GB10 and from $0.07\times$ to $0.88\times$ on H100, the largest value being that
        of $(192,128)$. Per iteration it is 4.5 to 26 times slower than the fixed-precision layer on
        GB10 and 3.6 to 19 times on H100, the ratio being largest at low precision, as in the AXPY
        comparison of Table~\ref{tab:cump}.
  \item \textbf{STS.} The \texttt{cu\_fcomplex} implementation is faster than the CPU only in 7 of the 42
        cases on GB10, all at $n\ge128$ and at most 256 bits (at best $1.5\times$), and in none on H100
        ($0.03$--$0.69\times$); the \texttt{cu\_mpc} implementation is faster in none on either machine
        ($0.02$--$0.33\times$ and $0.02$--$0.19\times$). In the \texttt{cu\_fcomplex} implementation one
        thread handles one root, so that at most $n\le192$ threads are active and each executes a long
        sequential computation. This is the situation of the thread-per-row matrix--vector product of
        Section~\ref{sec:blas}, and likewise the result characterizes this implementation rather than the
        method.
\end{itemize}
The two dependences have straightforward origins. The number of active threads grows with $n$ (the
Ehrlich--Aberth implementation uses $n$ blocks), so that the GPU is better utilized at high degree. The cost
of one multiplication grows quadratically with the precision on the GPU (schoolbook multiplication in both
layers), whereas the system GMP on the CPU uses assembly kernels and, at larger sizes, subquadratic
multiplication; the ratio therefore moves in favor of the CPU as the precision increases.

\paragraph{Comparison of the two machines}
The speedups are smaller on H100 for two reasons, both visible in the times per iteration. First, with at
most 192 blocks the GPU of H100 is not faster than that of GB10: the \texttt{cu\_fcomplex} implementations
need $1.4$--$1.7\times$ (Ehrlich--Aberth) and $1.5$--$1.9\times$ (STS) more time per iteration on H100 in
all 84 cases, and the \texttt{cu\_mpc} implementations between $0.63\times$ and $1.27\times$ the time on
GB10 (median $1.06\times$ for both methods). This agrees with the measurement of
Section~\ref{sec:analysis}, in which H100 was the slower machine as long as at most a few thousand threads
were active. Second, the 32-core host of H100 is the faster CPU for all but the smallest problems: its
fastest configuration needs between $0.43\times$ and $1.74\times$ the time of the fastest configuration of
GB10 (median $0.72\times$ for Ehrlich--Aberth and $0.66\times$ for STS) and is faster in 69 of the 84
cases.

\begin{figure}[htb]\centering
  \includegraphics[width=\linewidth]{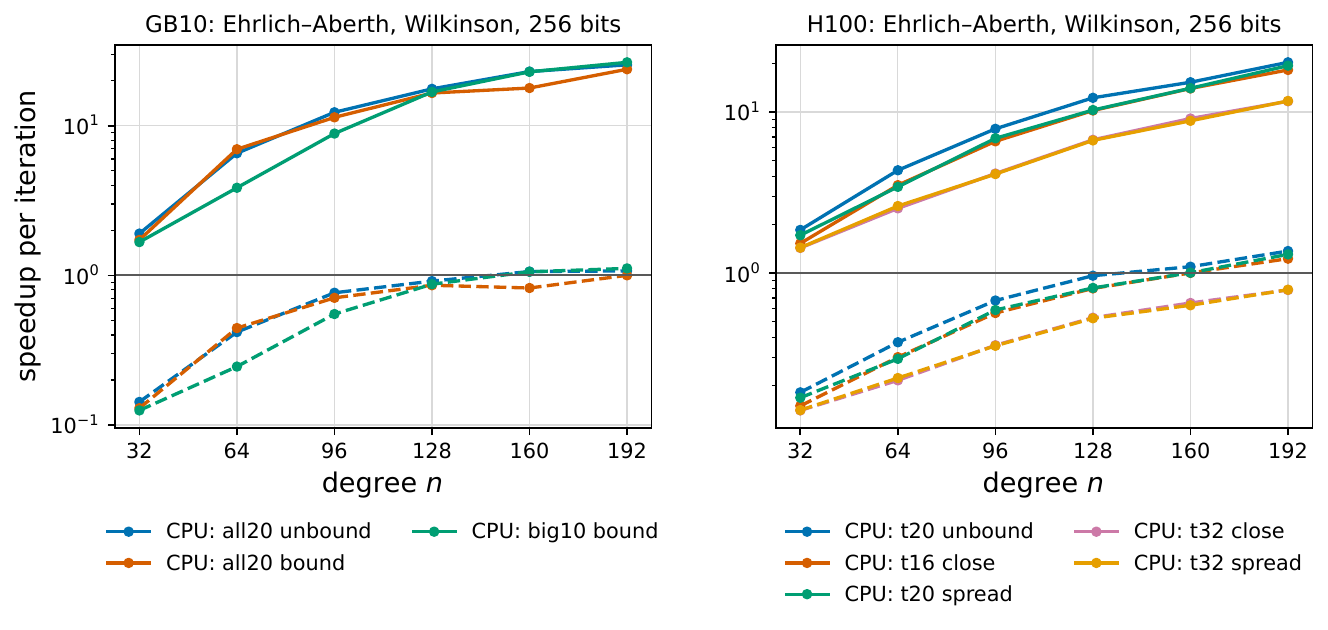}
  \caption{Higher-order Ehrlich--Aberth, Wilkinson polynomial, 256 bits: speedup per iteration against each
  CPU thread configuration (left: GB10, right: H100). Solid lines: \texttt{cu\_fcomplex}; dashed lines:
  \texttt{cu\_mpc}.}
  \label{fig:root_threads}
\end{figure}

\paragraph{Effect of the CPU configuration}
The choice of the CPU configuration matters. Over the 84 throughput cases on GB10, the fastest
configuration was the one with 10 threads on the Cortex-X925 cores in 55 cases, 20 bound threads in 22, and
20 unbound threads in 7; the slowest of the three configurations was up to $2.8\times$ slower than the
fastest (median $1.16\times$). On H100 the fastest configuration used 32 threads in 78 cases (\code{spread}
binding in 42, \code{close} in 36) and 20 unbound threads in the remaining 6, which are small problems;
the slowest of the five configurations was up to $3.0\times$ slower than the fastest (median
$1.96\times$). The configuration with 20 unbound threads corresponds to the one used throughout the
previous version of this paper; using it as the only baseline would have overstated the GPU speedups on both
machines. Figure~\ref{fig:root_threads} shows the speedups against each configuration separately for one
precision.

\subsection{Time to solution}
\label{sec:root_tts}

The time to solution is compared for the 19 cases per method in which the criterion is met. Since the
three implementations need the same number of iterations and attain comparable forward errors (within a
factor of 5.1, Section~\ref{sec:root_conv}), this is a comparison at comparable accuracy. Table~\ref{tab:tts} lists the
Ehrlich--Aberth results for $n\in\{32,96,192\}$; each time is the minimum of three runs.

\begin{table*}[t]
\centering
\caption{Time to solution [s] of the higher-order Ehrlich--Aberth method (wall-clock time until the
convergence criterion is met), number of iterations (identical for the three implementations), and largest
forward error among the three. CPU: fastest thread configuration. The last column of each machine is the
CPU time divided by the \texttt{cu\_fcomplex} time.}
\label{tab:tts}
\footnotesize\setlength{\tabcolsep}{4pt}
\begin{tabular}{lrrrl rrrr @{\hskip 2em} rrrr}
\toprule
 & & & & & \multicolumn{4}{c}{GB10} & \multicolumn{4}{c}{H100} \\
\cmidrule(lr){6-9}\cmidrule(lr){10-13}
problem & $n$ & bits & it & fwd.\ err. & CPU & \texttt{cu\_mpc} & \texttt{cu\_fc} & ratio & CPU & \texttt{cu\_mpc} & \texttt{cu\_fc} & ratio \\
\midrule
Wilkinson & 32 & 512 & 14 & 4.3\mbox{e}-131 & 0.007 & 0.054 & 0.0066 & 1.0 & 0.011 & 0.068 & 0.0108 & 1.0 \\
Wilkinson & 32 & 1024 & 14 & 1.7\mbox{e}-285 & 0.011 & 0.092 & 0.0172 & 0.6 & 0.014 & 0.116 & 0.0271 & 0.5 \\
Wilkinson & 96 & 1024 & 36 & 2.8\mbox{e}-236 & 0.270 & 0.578 & 0.0828 & 3.3 & 0.141 & 0.615 & 0.1261 & 1.1 \\
\midrule
Chebyshev & 32 & 256 & 9 & 5.5\mbox{e}-73 & 0.003 & 0.023 & 0.0021 & 1.5 & 0.005 & 0.028 & 0.0032 & 1.6 \\
Chebyshev & 32 & 512 & 9 & 6.9\mbox{e}-150 & 0.004 & 0.028 & 0.0040 & 0.9 & 0.006 & 0.035 & 0.0064 & 0.9 \\
Chebyshev & 32 & 1024 & 10 & 4.1\mbox{e}-304 & 0.007 & 0.049 & 0.0114 & 0.6 & 0.010 & 0.062 & 0.0177 & 0.5 \\
Chebyshev & 96 & 256 & 19 & 3.2\mbox{e}-61 & 0.048 & 0.098 & 0.0066 & 7.3 & 0.043 & 0.110 & 0.0100 & 4.2 \\
Chebyshev & 96 & 512 & 19 & 2.2\mbox{e}-138 & 0.078 & 0.125 & 0.0133 & 5.8 & 0.050 & 0.135 & 0.0216 & 2.3 \\
Chebyshev & 96 & 1024 & 20 & 3.1\mbox{e}-292 & 0.149 & 0.237 & 0.0369 & 4.0 & 0.070 & 0.227 & 0.0594 & 1.2 \\
Chebyshev & 192 & 512 & 35 & 4.0\mbox{e}-120 & 0.448 & 0.568 & 0.0381 & 11.8 & 0.359 & 0.470 & 0.0632 & 5.7 \\
Chebyshev & 192 & 1024 & 35 & 1.4\mbox{e}-274 & 0.779 & 1.174 & 0.1201 & 6.5 & 0.485 & 0.746 & 0.1791 & 2.7 \\
\bottomrule
\end{tabular}
\end{table*}

The \texttt{cu\_fcomplex} implementation of the Ehrlich--Aberth method obtains the roots faster than the
fastest CPU configuration in 16 of the 19 cases on GB10 and in 14 on H100. On GB10 it is
$1.6$--$11.8\times$ faster for $n\ge64$, the largest factor being that of the Chebyshev nodes with $n=192$
at 512 bits (0.038\,s against 0.45\,s, 35 iterations, forward error $4\times10^{-120}$); for $n=32$ the
ratio is $0.61$--$1.45\times$, the CPU being faster at 1024 bits and, for the Chebyshev problem, at 512
bits. On H100 it is
$1.1$--$5.7\times$ faster for $n\ge96$ (the same case being the best one: 0.063\,s against 0.36\,s); for
$n=64$ the ratio is $0.70$--$3.0\times$ and for $n=32$ $0.52$--$1.6\times$, the CPU being faster at 1024
bits for $n=64$ and, for $n=32$, in the same three cases as on GB10. The \texttt{cu\_mpc} implementation is slower than the
CPU in all 19 cases on both machines ($0.11$--$0.79\times$ on GB10, $0.12$--$0.76\times$ on H100). For the
STS method, both GPU implementations are slower than the CPU in all 19 cases: $0.10$--$0.56\times$ for
\texttt{cu\_fcomplex} and $0.02$--$0.33\times$ for \texttt{cu\_mpc} on GB10, and $0.04$--$0.20\times$ and
$0.02$--$0.18\times$ on H100, for the reason given in Section~\ref{sec:root_throughput}.

The per-iteration speedups of Section~\ref{sec:root_throughput} carry over to the time to solution because
the iteration counts are equal, but the largest of them do not correspond to solved problems: the
throughput ratios of $28$--$31\times$ (GB10) and $16\times$ (H100) are obtained at $n=192$ and 128 bits,
whereas already at 256 bits the criterion is not met for $n\ge128$ in either problem (the convergence
experiments do not include 128 bits). A speedup that represents this application is the factor in the time
to solution of the Chebyshev problem at $n\ge128$ and 512 to 1024 bits---$4.6$--$11.8\times$ on GB10 and
$1.7$--$5.7\times$ on H100---not the peak throughput ratio.

\section{Discussion: scope of applicability and limitations}
\label{sec:discussion}

\subsection{Which implementation to select}

Table~\ref{tab:guide} collects the measured results into a selection guide. It is based only on the
workloads and sizes measured in this paper. The recommendations are the same for the two machines; only
the factors differ, and both are given.

\begin{table*}[t]
\centering
\footnotesize
\caption{Selection guide derived from the measurements in this paper. Where two values are separated by a
slash, the first is for GB10 and the second for H100; ``all cores'' are the 20 and 32 CPU cores of the
respective machine.}
\label{tab:guide}
\begin{tabular}{@{}p{0.30\textwidth}p{0.13\textwidth}p{0.50\textwidth}@{}}
\toprule
situation & recommended & evidence \\
\midrule
Precision known at compile time, up to 2048 bits; correctly rounded $+,-,\times$ suffice, and division
within the tested accuracy is acceptable; thousands of independent work items
& \freal/\fcomplex
& Fastest of the four implementations in AXPY up to 2048 bits on both machines (Table~\ref{tab:cump});
  Ehrlich--Aberth iteration faster than the fastest CPU configuration in 39\,/\,36 of 42 cases, by up to
  $31\times$\,/\,$16\times$; the exceptions are at $n=32$ with 512 or 1024 bits and, on H100, at $n=64$
  with 1024 bits
  (Table~\ref{tab:root_corner}); time to solution $1.6$--$11.8\times$ shorter for $n\ge64$\,/\,
  $1.1$--$5.7\times$ shorter for $n\ge96$ (Table~\ref{tab:tts}) \\ \addlinespace
Precision of 4096 bits or more; only $+,-,\times$; no rounding guarantee needed
& CUMP
& Fastest in AXPY from 4096 bits on both machines with the default settings of the released header
  (Table~\ref{tab:cump}); with the truncated product enabled up to 128 limbs, \freal{} is faster than CUMP
  at 4096 and 8192 bits on GB10 but not on H100 (Section~\ref{sec:cump}) \\ \addlinespace
MPFR/MPC semantics required \emph{inside} a kernel: directed rounding, ternary value and flags, exponent
range, run-time precision, or correctly rounded elementary functions
& \code{cu\_mpfr}/\code{cu\_mpc}
& The only GPU implementation providing these semantics (Tables~\ref{tab:libs} and~\ref{tab:verify}),
  including the upstream behavior of Appendix~\ref{sec:appendix_tie}.
  Cost: AXPY 3.6--11$\times$\,/\,8.7--15$\times$ slower than CUMP (Table~\ref{tab:cump}). Against the
  system libraries on all cores at 1024 bits: AXPY 2.2--6.9$\times$\,/\,4.2--5.9$\times$ faster,
  block-per-row matrix--vector product 1.1--2.2$\times$\,/\,1.9--6.4$\times$ faster, matrix--matrix
  product 0.96--1.24$\times$\,/\,0.66--2.5$\times$, elementary functions not faster except complex
  \code{sqr} (Tables~\ref{tab:elem_real}--\ref{tab:linalg}) \\ \addlinespace
Elementary functions of a few thousand arguments; algorithms with fewer than a few hundred independent
work items (thread-per-row matrix--vector product, one thread per root); or results that can be computed
on the host
& CPU (system MPFR/MPC)
& Run-time-precision layer 0.05--0.80$\times$\,/\,0.10--1.00$\times$ the speed of all cores for real
  functions at 1024 bits (Table~\ref{tab:elem_real}); thread-per-row matrix--vector product
  0.12--0.68$\times$\,/\,0.07--0.62$\times$ (Table~\ref{tab:linalg}); STS with one thread per root below
  $1\times$ in 35\,/\,42 of 42 cases (Section~\ref{sec:root_throughput}) \\
\bottomrule
\end{tabular}
\end{table*}

Two factors are common to these results. The first is the number of independent multiple-precision
operations available at one time. A single GPU thread executes the run-time-precision layer two orders of
magnitude more slowly than a CPU core executes the system MPFR (Section~\ref{sec:analysis}); the GPU is
competitive only when thousands of such computations run concurrently. Both layers should therefore be
used with algorithms formulated to expose this parallelism (one block per row or per root, tree
reductions), and algorithms whose parallelism is bounded by a small problem dimension are better served by
the CPU. The second factor is the code that each thread executes. The fixed-precision layer, whose
kernel is compiled without register spills up to 1024 bits and which performs no allocation, is 9--28$\times$ (GB10) and
14--24$\times$ (H100) faster than the run-time-precision layer in AXPY up to 1024 bits and is the only one of the two that outperforms a
multithreaded CPU by a wide margin in the application. The run-time-precision layer should accordingly be
understood as a means of making the functionality and the semantics of MPFR/MPC available where the data
are, on the device, and not as a general-purpose route to acceleration; where a computation needs only
correctly rounded $+,-,\times$, the fixed-precision layer is the appropriate choice.

\subsection{Limitations}

\begin{itemize}
  \item \textbf{Speed relative to the system libraries.} The run-time-precision layer is built on
        mini-GMP, which on a CPU core is between 2.7 and 44 times (GB10) or 2.3 and 37 times (H100) slower
        than the GMP-based system MPFR/MPC for the functions measured
        (Section~\ref{sec:elementary_function}). On neither machine does the GPU recover this factor for
        the elementary functions relative to all CPU cores; for the matrix--matrix product it recovers it
        only marginally on GB10 and by a factor of at most 2.5 on H100. Subquadratic multiplication and division are not implemented in either
        layer.
  \item \textbf{Speed of the run-time-precision layer relative to other GPU libraries.} Its AXPY is
        3.6--11$\times$ (GB10) and 8.7--15$\times$ (H100) slower than that of CUMP and up to 28$\times$
        and 26$\times$ slower than that of the fixed-precision layer (Section~\ref{sec:cump}), because every operation goes through the general
        MPFR code path with temporaries in device memory.
  \item \textbf{High precision in the fixed-precision layer.} Above 1024 bits the compiler reports register
        spills, and with the default code path the time of a multiplication increases abruptly
        above 2048 bits (Section~\ref{sec:cump}).
  \item \textbf{Guarantees of the fixed-precision layer.} Only round-to-nearest is provided; there are no
        infinities, NaNs, signed zeros, exception flags, or exponent range checks; and only addition,
        subtraction, and multiplication are correctly rounded. Division, square root, and the functions
        of the first group of Table~\ref{tab:fixed_ulp} were within 1\,ulp in our tests without a proof;
        the trigonometric, inverse trigonometric, and complex functions lose accuracy without bound near
        the arguments listed in Section~\ref{sec:verification} and must not be used there.
  \item \textbf{Verified but not proved.} The agreement of the run-time-precision layer with MPFR/MPC is
        established by tests (Table~\ref{tab:verify}), not by a proof that the conversion preserves the
        semantics, and the conversion did contain defects that only the full-significand tests exposed
        (Section~\ref{sec:div}). The converted functions that do not appear in Table~\ref{tab:verify}
        were not tested to the same extent, and the tests were run with a single compiler version.
  \item \textbf{Per-thread state.} The exponent range and the flags are separate for each GPU thread only
        after \code{mpc\_cuda\_tstate\_install}; without it they are shared and concurrent use is
        incorrect (Table~\ref{tab:shared}). The default precision and the default rounding mode are also
        kept per thread, but the caches of constants are bypassed on the device, so that constants such
        as $\pi$ and $\log 2$ are recomputed in every call that needs them.
  \item \textbf{Functionality.} The special functions of MPFR, formatted I/O on the device, and the
        \code{mpf}/\code{mpq} conversions are not available.
  \item \textbf{Scope of the comparison.} The comparison with other GPU libraries is limited to real AXPY
        with CUMP and CAMPARY and to AXPY and matrix products with MPRES-BLAS, whose precisions do not
        coincide with ours (Section~\ref{sec:mpres}). GARPREC could not be built, and the libraries
        specialized for at most 212 bits were not measured; no claim is made about them.
  \item \textbf{Machines and workloads.} The measurements cover two machines, one compiler version, and
        problem sizes up to a few times $10^4$ concurrent work items. The qualitative statements of
        Table~\ref{tab:guide} hold on both machines, but the factors differ by up to several times, and in
        the root-finding application the larger GPU was the slower one for the fixed-precision
        implementations, in which at most 192 blocks are active. Extrapolation to other machines or to much larger problems is not supported by the data.
        No hardware performance counters were available on either machine, so that the explanations given
        for the measured times in Section~\ref{sec:analysis} rest on timing experiments only. On H100,
        MPRES-BLAS failed intermittently for the larger moduli sets, which restricts that comparison
        (Section~\ref{sec:mpres}).
\end{itemize}

\section{Conclusion}

We presented mpc\_cuda, an arbitrary-precision arithmetic environment for CUDA that consists of a
script-converted port of the mini-GMP-based MPFR and MPC, with per-thread memory management and per-thread
MPFR state, and of compile-time fixed-precision real and complex types based on the same limb-array
representation.

The findings are as follows; where two values are given, the first is for GB10 and the second for H100.
(1)~\emph{Functionality and reliability.} The functionality of MPFR/MPC---run-time precision, all rounding
modes, ternary values, flags, exponent range, and correctly rounded elementary functions for real and
complex arguments---is available inside CUDA kernels. In 3.3 million test cases executed concurrently on
the GPU, no discrepancy from the system libraries was found on either machine: every result, ternary
value, and flag was identical in the 3.02 million cases with deterministic rounding modes, and every value
was admissible in the 0.28 million cases with faithful rounding. This agreement was obtained only after the same tests had exposed two defects of the conversion
and a data race on MPFR's global state, none of which the binary64-based tests of the previous version
had detected; a port of this kind should not be trusted without a comparison over the full significand.
(2)~\emph{Performance of the run-time-precision layer.} Executing the ported code on the GPU is
17--31$\times$\,/\,14--77$\times$ faster than executing it on one CPU core for elementary functions at
1024 bits and 54--70$\times$\,/\,53--193$\times$ faster for AXPY, the block-per-row matrix--vector product,
and the matrix--matrix product, provided that thousands of threads are active. Because the mini-GMP-based
code is itself 2.7--44$\times$\,/\,2.3--37$\times$ slower than the GMP-based system libraries, the gain over
the system libraries on all CPU cores is much smaller: 2.2--6.9$\times$\,/\,4.2--5.9$\times$ for AXPY,
1.1--2.2$\times$\,/\,1.9--6.4$\times$ for a matrix--vector product that assigns one block per row, and
0.96--1.24$\times$\,/\,0.66--2.5$\times$ for the matrix--matrix product; the elementary functions are not
faster on either machine, with the exception of the complex square. The purpose of this layer is to provide
MPFR/MPC semantics on the device, not to accelerate them.
(3)~\emph{Performance of the fixed-precision layer.} For AXPY, the fixed-precision type is
9--28$\times$\,/\,14--24$\times$ faster than the run-time-precision layer up to 1024 bits, faster than
CAMPARY at all precisions from 128 to 8192 bits, and faster than CUMP up to 2048 bits; from 4096 bits CUMP
is faster on both machines with the default settings of the header. Only its addition, subtraction, and multiplication are correctly rounded.
(4)~\emph{Application.} In simultaneous polynomial root-finding with the higher-order Ehrlich--Aberth
method, the three implementations, given identical rounded coefficients and initial values, need
identical numbers of iterations in all cases that converge and attain forward errors within a factor of
5.1 of each other, on both machines. At this comparable
accuracy the fixed-precision implementation obtains the roots 1.6--11.8$\times$ faster than the fastest
multithreaded CPU configuration for degrees 64 to 192 on GB10 and 1.1--5.7$\times$ faster for degrees 96 to
192 on H100; at lower degrees it is not consistently faster, and the run-time-precision implementation is
slower than the CPU in all cases. (5)~\emph{Machines.} The larger GPU is the faster one only when enough
threads are active: in the linear algebra kernels that launch 4096 to 32768 threads H100 is
1.3--4.3$\times$ faster than GB10, whereas in the thread-per-row matrix--vector product (at most 1024
threads) and in the fixed-precision root-finding programs (at most 192 blocks) GB10 is the faster machine
by 16\% to 91\%. Together, these results delimit the
regime in which each layer should be used (Table~\ref{tab:guide}).

Future work includes replacing mini-GMP by a multiplication and division layer designed for the GPU,
including subquadratic algorithms, a memory layout that allows coalesced access in the run-time-precision
layer, argument reduction with rigorous error bounds for the fixed-precision functions, and the
application of both layers to iterative solvers for linear systems such as GMRES($m$) with preconditioning
on ill-conditioned problems from the SuiteSparse matrix collection~\cite{sparse_matrix_collection}.

\section*{Acknowledgments}
The author thanks the anonymous reviewers for their careful reading and valuable comments, which led to
substantial improvements of this paper.
This work was supported by JSPS KAKENHI Grant Number JP26K14846.

\appendix
\section{Use of a generative AI assistant in development}
\label{sec:appendix_ai}

The conversion scripts and the fixed-precision headers were developed with Claude Code~\cite{claude_code}, a
generative AI coding assistant, under the direction of the author. Development proceeded in five stages
(mini-GMP, MPFR basic operations, MPFR transcendental functions, MPC, and an AXPY demonstration), and the
results of each stage were compared with the host before proceeding to the next. The assistant was most
useful in diagnosing device-side failures (infinities, meaningless values, illegal memory accesses) by
reducing them to small reproducing programs; an undersized device stack was found in this way. The
same procedure had led to the replacement of \code{mpfr\_div} in the previous version, which the
verification of this paper showed to be unnecessary (Section~\ref{sec:div}); the benchmark and verification
programs of this revision, and the corrections of the conversion scripts reported in
Section~\ref{sec:div}, were also written with the assistant. We do not claim the
development procedure as a contribution, and all numerical claims in this paper rest on the tests described
in Section~\ref{sec:verification}, not on the provenance of the code.

\section{An exact midpoint for which \code{mpfr\_div} does not round to even}
\label{sec:appendix_tie}

The following instance is one of those referred to in Section~\ref{sec:verification}. Let
{\small
\begin{align*}
u &= \mathtt{0x13714e8d132ad030a2ed38}\\[-2pt] &\phantom{{}=\mathtt{0x}}\mathtt{b3d4bff0c96038bc4b87582},\\
v &= \mathtt{0x1782a2d35dad808a9ef1e5877dfba},
\end{align*}}
integers of 177 and 113 bits. Then $u=v\cdot c$ exactly with the odd 64-bit integer
$c=\mathtt{0xd3b5207c4011ba35}$, so that $u/v=c$ lies exactly halfway between the two neighboring 63-bit
numbers $c-1$ and $c+1$. The 63-bit significand of $c-1$ is even and that of $c+1$ is odd; round-to-nearest-even
must return $c-1=\mathtt{0x\ldots ba34}$. With $u$ and $v$ stored exactly in variables of 177 and 113 bits,
\code{mpfr\_div} with a 63-bit result and \code{MPFR\_RNDN} returns $c+1=\mathtt{0x\ldots ba36}$ with a positive
ternary value. We observed this with MPFR~4.2.2 without patches (\code{mpfr\_get\_patches} returns an empty
string) on GMP~6.3.0 with 64-bit limbs, both built from source with their default configuration on GB10
(AArch64, GCC~13.3.0), and also with the MPFR~4.2.1 package of Ubuntu~24.04 on the same machine. The program
\code{verify/mpfr\_div\_tie\_repro.c} in the distributed experiments generates such cases ($u=v\cdot c$ with
random $v$ and random odd $c$ of one bit more than the precision of the quotient), links only GMP and MPFR,
and checks each quotient against the value obtained by integer arithmetic.

The cause is a missing case in the general division routine (\code{src/div.c}). When the divisor is
truncated and the estimated quotient $q_1$ is one unit of its last limb too large, round-to-nearest returns
$q_1$ whenever the round bit and the remaining low bits of $q_1$ are zero. This is correct unless the
precision is $63 \bmod 64$ bits and $u/v=q_1-1$ exactly: $q_1-1$ is then the midpoint of the representable
neighbors $q_1-2$ and $q_1$, and the even rule must select $q_1-2$ when the last significand bit of $q_1$
is one, as in the instance above ($q_1=c+1$). With a change of that branch that applies the even rule in
this case, a comparison of \code{mpfr\_div} with quotients and ternary signs computed by integer arithmetic
(189 million divisions in the five rounding modes, with precisions of 2 to 319 bits and divisors of 2 to
1000 bits) shows no difference, whereas the unpatched library returns 200,638 wrong values, all of them
for exact midpoints, round-to-nearest, and precisions of $63 \bmod 64$ bits; the test suite of MPFR passes
in both cases. We reported the observation, the analysis, and the proposed change to the MPFR developers
through the MPFR mailing list on December 1, 2026. Until a corrected version is released, results of
\code{mpfr\_div}, and hence of \code{cu\_mpfr\_div}, can differ from round-to-nearest-even in these
cases.

\section{Reproducibility information}
\label{sec:appendix_repro}

The programs, scripts, and raw results of Sections~\ref{sec:verification}--\ref{sec:rootfinding} that use
only mpc\_cuda and the system GMP, MPFR, and MPC are distributed with the library in the directory
\code{experiments/} of the public repository~\cite{mpc_cuda} (branch \code{v0.0.2-dev}): the script
\code{run\_all.sh} builds the library and runs these measurements serially (one program at a time, with
nothing else running on the machine), and \code{make\_paper\_tables.py} generates the tables of this paper
from the resulting CSV files. The programs that use the other libraries (the CUMP and CAMPARY versions of
the AXPY benchmark of Section~\ref{sec:cump} and the MPRES-BLAS benchmark of Section~\ref{sec:mpres}) and
the sources of these libraries are not distributed; for them, only the measured results are included, and
the versions and changes needed to rebuild them are listed below. The experiments were run with mpc\_cuda
at commit \code{83c78d6} of that repository.

\begin{description}
  \item[Upstream sources] mini-GMP from GMP~6.3.0, MPFR~4.2.2, MPC~1.4.1.
  \item[System libraries] (CPU baselines and references) GMP~6.3.0 and MPFR~4.2.2 built from source with
        their default configuration (assembly kernels enabled, thread-safe; on H100 GMP was configured
        in addition with \code{-{}-enable-cxx}), and MPC~1.4.1 built against them by the benchmark scripts
        and linked statically into the verification, kernel, and root-finding programs. The MPC installed
        system-wide is not used; on GB10 it is version~1.4.0, and with it the CPU root-finding program
        behaves differently for symmetric initial values (Section~\ref{sec:root_conv}).
  \item[Compared libraries] CUMP~1.0.1 (commit \code{8e81ae5}) with two local changes that replace CUDA
        intrinsics and runtime calls removed from current CUDA versions (\code{\_\_any}/\code{\_\_all},
        \code{cudaThreadSynchronize}) and do not touch the arithmetic; CAMPARY \code{campary\_01.06.17}
        (\code{Doubles/src\_gpu}, \code{multi\_prec<K>} with the certified algorithms); MPRES-BLAS~1.7.0
        (commit \code{427ba19}), compiled with the additional option \code{-include cstdint}.
  \item[Compilers and options] CUDA~13.0 (nvcc~13.0.88), GCC~13.3.0; \code{sm\_121} (GB10) and
        \code{sm\_90} (H100). The library is compiled with \code{-fmad=false -DNO\_ASM} and nvcc's default
        optimization of the device code; the host half of the library, which is the host path used as a
        baseline, is compiled with \code{-O2} (nvcc passes no optimization option to the host compiler
        unless one is given; the previous version was therefore compiled without host optimization). The
        kernel benchmarks of Sections~\ref{sec:elementary_function}--\ref{sec:analysis} use
        \code{-fmad=false}; the AXPY comparison uses \code{-O3 -fmad=false} for all four libraries; the
        root-finding programs use \code{-O2 -fmad=false -std=c++17}. The CPU baselines are compiled with
        \code{g++ -O2 -fopenmp}.
  \item[Device limits] \code{cudaLimitStackSize} is 256\,KB for the verification, the kernel benchmarks,
        and the root-finding programs, and 64\,KB for the \code{cu\_mpfr} AXPY of Section~\ref{sec:cump}.
        \code{cudaLimitMallocHeapSize} is 1\,GB for the verification and the root-finding programs,
        2\,GB for the kernel benchmarks (it is used only in the \code{malloc} rows of
        Table~\ref{tab:arena}), and 64\,MB for the AXPY comparison.
  \item[Launch and arena sizes] Verification: $32\times256$ threads, 2\,MB slabs. Elementary functions:
        $256\times32$ threads, 1\,MB (real) and 2\,MB (complex) slabs. Real AXPY: $256\times64$ threads,
        32\,KB slabs; complex AXPY: $256\times32$ threads, 256\,KB slabs. Thread-per-row matrix--vector
        product: $N$ threads; block-per-row: $N$ blocks of 32 threads; matrix--matrix product: one thread
        per element, at most 16384 threads. AXPY comparison: see Section~\ref{sec:cump}. Root-finding:
        one block per root for the Ehrlich--Aberth method and for the \code{cu\_mpc} version of the STS
        method; one thread per root for the \fcomplex{} version of the STS method.
  \item[CPU baselines] Sections~\ref{sec:elementary_function}--\ref{sec:blas}: system MPFR/MPC with one
        thread and with one OpenMP thread per core (\code{OMP\_PROC\_BIND=close}, \code{OMP\_PLACES=cores};
        20 threads on GB10, 32 on H100), mean of three runs; host path of mpc\_cuda with one thread, one
        run. The timed region contains only the loop over the work items. Section~\ref{sec:rootfinding}:
        system MPFR/MPC with the thread configurations listed there, the fastest being used for each case.
  \item[Timing] GPU: \code{cudaEventRecord} around the kernel launch after one warm-up launch; mean of 10
        launches (20 in Section~\ref{sec:cump}; 3 for fewer than 64 launched threads in
        Table~\ref{tab:threads}). Root-finding: the whole iteration loop, including the transfer of the $n$
        corrections to the host in each iteration.
  \item[Inputs] Verification: Section~\ref{sec:verification}, seed 20261130. Kernel benchmarks: uniformly
        distributed random numbers with full \PB-bit significands generated by MPFR
        (\code{mpfr\_urandomb} scaled to the stated interval) with fixed seeds; vectors and matrices in
        $[0,1)$. AXPY comparison: random \PB-bit values in $[0,1)$ from GMP's generator with a common seed;
        reference at $\PB+64$ bits. Root-finding: Section~\ref{sec:root_problems}.
\end{description}


\end{document}